\documentclass[12pt]{article}
\usepackage{graphicx}
\usepackage{float}
\usepackage{amsmath}
\usepackage{cases}
\usepackage{amsfonts}
\usepackage{stfloats}
\usepackage{cite}
\usepackage{arydshln}
\usepackage{multirow}
\usepackage[top=1in, bottom=1in, left=1in, right=1in]{geometry}

\begin{document}
\title{Performance Analysis of Slotted Secondary Transmission with Adaptive Modulation under Interweave Cognitive Radio Implementation}
\author{
Wen-Jing Wang and Hong-Chuan Yang}
\date{}
\maketitle
\begin{abstract}
In cognitive radio communication, unlicensed secondary user (SU) can access under-utilized spectrum of the licensed primary user (PU) opportunistically for emerging wireless applications. With interweave implementation, SU has to perform spectrum sensing on the target frequency band and waits for transmission if PU occupies the channel. This waiting time results in extra delay for secondary transmission. In this paper, the delay and throughput performance of secondary packet transmission is evaluated with slotted transmission protocol. We propose a discrete-time Markov model to characterize secondary slotted transmission process. Close-form solution of collision probability is obtained. We then carry out the queuing delay and throughput analysis based on a two-dimensional-finite-state Markov chain for small-size packet transmission. For large-size packets, the distribution function of extended delivery time for secondary packet transmission is also derived. Selected numerical results are presented to illustrate the mathematical formulas and to validate our research results.
\end{abstract}

\section{Introduction}
Cognitive radio system is a promising solution to solve radio spectrum scarcity problem. Secondary user (SU) can explore the under-utilized licensed frequency bands of primary system with opportunistic spectrum access (OSA) strategies. On the other hand, interference caused by secondary system has to be properly controlled so that the primary user's (PU) communication is not significantly affected by opportunistic channel access. Various implementation strategies can apply to achieving opportunistic spectrum sharing \cite{788210,1391031,1000009}. With underlay cognitive strategy, SU and PU can transmit simultaneously subject to a SU-to-PU interference constrain. As such, SU transmitter has to acquire SU-to-PU channel information, which can be very challenging in practice. With interweave implementation, SU can transmit only when PU is off and must vacate the spectrum when PU starts transmission. Hence, SU transmitter has to monitor PU activity on the target frequency band and perform spectrum hand-off adaptively for relinquishing or reaccessing the channel. Typically, multiple spectrum hand-off are involved to complete the secondary transmission of a given amount of data, resulting in extra delay. The total service time consists an interleaved sequence of transmission slots and waiting slots.

With interweave implementation, SU performs spectrum sensing\footnote{In this work, we focus on perfect sensing scenario, ongoing effort is being carried out to extend the analysis to imperfect sensing.} for transmission availability. Ideally, SU continuously monitors the target spectrum while transmitting and waiting \cite{4481339}. As such, no interference is introduced, at the cost of higher energy consumption and implementation complexity. Alternatively, SU can also perform spectrum sensing on a periodic basis. In \cite{7111368}, the statistics of extended delivery time (EDT) is evaluated with continuous spectrum sensing and semi-periodic spectrum sensing strategy, where SU continuously senses the target spectrum while transmitting and periodically senses for availability while waiting. Assuming continuous spectrum sensing, \cite{wang_ted} evaluates the statistics of EDT of SU packet transmission under interweave fashion with adaptive modulation. Substantial amount of previous work has been carried out in developing practical accessing strategies with various PU and SU models. Assuming that all the primary users and secondary user share the same slotted transmission structure, authors in \cite{4155374} design a novel MAC protocol based on optimal and suboptimal spectrum sensing and accessing strategies, where the optimization problem is formulated as a partially observable Markov Decision Process. \cite{4410464} proposes optimal and suboptimal access protocol by maximizing the throughput of secondary transmission subject to the collision probability constrain, where secondary user adopts slotted transmission coupled with periodic sensing strategy. Authors in \cite{6104049} propose an optimal sensing order selection strategy for multiple PUs' and multiple SUs' cognitive radio network, where the proposed sensing strategy converges to the collision free channel sensing order when the number of SUs' is less than the number of the channels. In this work, we consider a single PU and single SU cognitive communication scenario, where SU adopts slotted transmission strategy. Collision occurs with periodic spectrum sensing when PU starts transmission, however, SU has not sensed the channel yet. Different from previous works, we focus on collision, delay and throughput analysis of SU with slotted communication protocol.
%Spectrum sensing is performed before each transmission period as secondary user transmits with a slotted structure. Thus, sensing error could impact the performance of secondary link. The analytical formulations of the throughput of cognitive multi-channel MACs with perfect and imperfect sensing are presented in \cite{5963087}, where a discrete time Markov chain is used to model the number of communicating node pairs in the MAC protocols. The authors in \cite{6678828} precisely analyzed the effect of imperfect sensing on the stability region of the multi-user OSA system. The condition with sensing errors, for which the identical stability region achieved with perfect sensing, is also presented. The effects of imperfect channel sensing decisions, interference from the primary user using Gaussian mixture model, are studied in \cite{6733249}. Joint algorithm for sensing adaptation and opportunistic resource allocation is proposed in \cite{6152075} to minimize the total expected cost of the throughput losses and utilization of unused frequency  bands.

The delay and throughput analysis for secondary transmission is of considerable current research interest in cognitive radio. \cite{6220294} investigates the average waiting time and average service time of SU in one transmission slot with general primary model. \cite{5502787} derives the distribution function of service time to SU within a fixed period of time. So as to evaluate the delay performance for secondary users, \cite{6167447} proposes a priority virtual queue model. In cooperative wireless communication, \cite{5956451} investigates the probability of successful transmission with the hard delay constraints. \cite{6549030} analyzes the end-to-end performance of an interweave cognitive radio network in terms of the throughput and delay. In \cite{5577709}, a queueing performance analysis is carried out for dynamic spectrum accessing of secondary users. A dynamic spectrum selection method is proposed in \cite{6805863} to minimize the delay for secondary transmission in a pre-emptive resume M/G/1 queueing network. SU delay and throughput evaluation with slotted transmission is barely investigated to the best of our knowledge.

The concept of EDT was first introduced to derive the throughput bounds and delay performance of SU in cognitive radio transmission system \cite{Ted}. Generally, EDT consists of an interleaved sequence of transmission slots and waiting slots. By taking into account of the waiting time during secondary packet transmission, EDT is a significant performance metric for cognitive systems. \cite{5738219} investigates EDT considering the spectrum sensing error.\cite{6331691} studies EDT for cognitive radio system with multiple available channels and multiple SUs. The statistical characteristic of EDT depends on both spectrum sharing strategy and packet transmission policy. SU can adopt either work-preserving strategy \cite{5738219,6331691}, where SU restarts transmission from the breaking point without wasting previous transmission, or non-work-preserving strategy, where the SU retransmits the whole packet after reaccessing to the channel. Work-preserving strategy is achievable with the help of rateless codes \cite{1561992,1576565,4381419}, and also applies to the transmission of individually-coded small packets. In \cite{7111368}, the exact PDF of EDT for secondary packet transmission is derived for work-preserving strategy. Corresponding queuing delay is also evaluated with the help of first-order and second-order moments of EDT. Here, we focus on work-preserving strategy.

Previous works on throughput and EDT investigation concentrate on waiting time analysis while assuming the transmission time is constant. Transmission time was evaluated as the ratio of packet size over instantaneous channel capacity for slow fading case or as the ratio of packet size over ergodic channel capacity for fast fading case. However, with adaptive modulation (AM), which can guarantee reliability, the information rate during transmission may be adjusted as packet transmission experiences various channel realization, leading to varying transmission time for fixed-length packet. AM can take advantage of better channel conditions to improve data throughput, with a guaranteed bit error rate (BER) \cite{950343}. SU can only access the channel when PU is off, throughput improvement is critical in secondary system design. The queuing performance with adaptive modulation and coding technique in conventional communication system is studied in \cite{1427704}. However, with cognitive implementation, the analytical approach is different since SU has to wait for transmission. %In this paper, we assume that secondary user adopts adaptive modulation and slotted communication structure. Specifically, we propose a two-dimensional Markov chain to evaluate the delay and throughput performance of secondary system. We also derive the PMF of EDT assuming fixed-rate transmission or adaptive modulation.

In this paper, we extend previous work on performance evaluation of secondary transmission by considering AM and practical sensing strategy. Specifically, we investigate the queuing performance for short-packet transmission and obtain the probability mass function (PMF) of EDT for long-packet transmission. The major contribution of this work can be summarized as:
\begin{enumerate}
\item We propose a three-state discrete-time Markov chain to model the secondary transmission with slotted communication protocol. With practical spectrum sensing strategy, secondary transmission may cause interference to primary communication. Thus, the steady-state probability is calculated to evaluate the collision probability.
\item We propose a two-dimensional Markov model, characterizing SU transmission process, secondary channel realization and queue dynamics, to evaluate the queuing delay, throughput and packet drop probability of short-packet transmission with AM.
\item For large-size packet, multiple slots are required to complete transmission. The analytical expression of PMF for EDT is also derived considering both fixed-rate transmission and AM. Selected numerical results are presented to illustrate and validate our analysis.
\end{enumerate} 

The rest of this paper is organized as follows. In Section II, we illustrate the system model, channel model, and problem formulation. In Section III, we introduce a three-state Markov chain to model slotted secondary transmission. In section IV, we study the secondary queuing performance with the help of a two-dimensional Markov model. The PMF of EDT for large packet transmission over secondary link is derived in Section V. The paper is concluded in Section VI.  

\section{System and channel Model}
We assume that SU opportunistically accesses a PU channel in an interweave fashion. The PU behaviour on the channel is modeled by a two-state homogeneous  continuous-time Markov chain, where state $B$ represents that PU is transmitting (i.e. channel unavailable for secondary transmission) and state $I$ corresponds that PU is not transmitting. Researches on IEEE 802.11 Wireless LAN (WLAN) support a semi-Markov model for various traffic types. As such, Markov model serves as a reasonable alternative \cite{4086336,markov,4197717}. The sojourn time in states $B$ and $I$ follow independent non-identical (i.n.d) exponential distribution with average $\lambda$ and $\mu$, respectively.

\begin{figure}[t]
\begin{center}
\includegraphics[height=5.5cm,width=8.8cm]{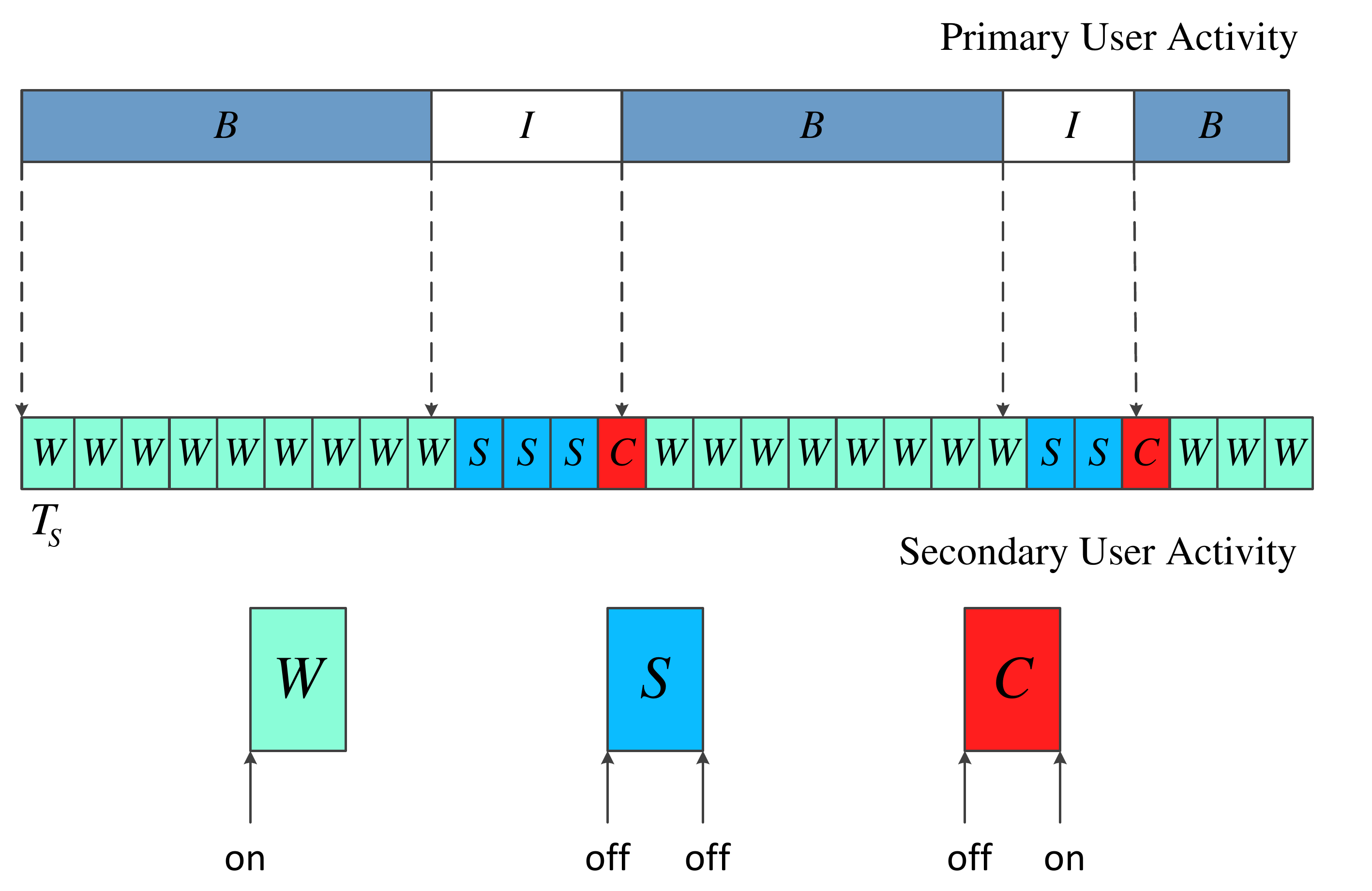}
\caption{Illustration of secondary slotted transmission with periodic spectrum sensing strategy.}
\label{FPS}
\end{center}
\end{figure}

SU employs a slotted transmission strategy, as illustrated in Fig. \ref{FPS}. SU periodically senses the PU channel with a fixed sensing interval $T_s$. SU transmission decision is based on sensing results. If PU channel is sensed busy, SU will wait for another $T_s$ duration and sense again. If the channel is sensed free, SU starts or continues its transmission immediately. Here, we assume that $T_s$ is sufficiently small, and the chance that PU activity changes twice in one $T_s$ is negligible. As such, if the channel is sensed free at two consecutive sensing instants, secondary transmission after the first sensing instant will not collide with PU transmission. However, if the channel is sensed busy after being sensed free, the secondary transmission will interfere primary transmission. We also assume that sensing duration is negligible compared to $T_s$. Hence, $T_s$ is approximately equal to the time duration that SU transmits or waits.

Secondary transmission adopts channel adaptive transmission over each transmission interval $T_s$. To implement adaptive modulation, SU transmitter sends a pilot signal once the spectrum sensed free.  The receiver estimates the received SNR from the pilot signal and determines which SNR region it falls into. Then, the receiver feeds back the index of the selected modulation scheme to the transmitter via an error-free feedback channel. After that, the transmitter and the receiver will be configured to use identical modulation scheme. Specifically, the transmission rate measured in $\text{bits/symbol}$ is chosen based on the secondary channel quality in each $T_s$. The value range of received signal-to-noise ratio (SNR) of secondary channel, denoted by $\gamma$, is divided into $N+1$ regions, $A_j,\ j=0,1,\cdots,N$. If $\gamma$ falls in region $A_j$ over a transmission interval $T_s$, rate $R_j\ \text{bits/symbol}$ will be used. The transmission rate remains constant over one $T_s$ and may change to a different value independently for next transmission interval.

The boundaries for SNR regions, $\gamma_j$'s, are selected to maintain a target instantaneous bit error probability $\text{BER}_{\textrm{tar}}$. For example, if we adopt $2^n$-ary square QAM modulation schemes, whose approximate BER is given by
\begin{equation}
\text{BER}_n(\gamma)=\frac{1}{5}\exp\left[-\frac{3\gamma}{2(2^n-1)}\right],\ \ n=1,2,\cdots ,N,\label{BERn}
\end{equation}
then, the boundary SNR can be calculated for $\text{BER}_{\textrm{tar}}$ as \cite{634685}
\begin{equation}
\gamma_j=-\frac{2}{3}\ln(5\text{BER}_{\textrm{tar}})(2^{j+1}-1),\ \ j=1,2,\cdots ,N.\label{gammaj}
\end{equation}
The probability that transmission rate $R_j$ is used, denoted by $\pi_j$, equals to the probability that $\gamma$ falls into region $A_j$, which can be calculated as
\begin{equation}
\pi_j=\int_{\gamma_{j-1}}^{\gamma_j}f_{\gamma}(\gamma)d\gamma, \label{Pro_Rj}
\end{equation}
where $f_{\gamma}(\gamma)$ represents the probability density function of received SNR. 

\section{Markov model for secondary transmission}
\begin{figure}[t]
\begin{center}
\includegraphics[height=4.5cm,width=8.8cm]{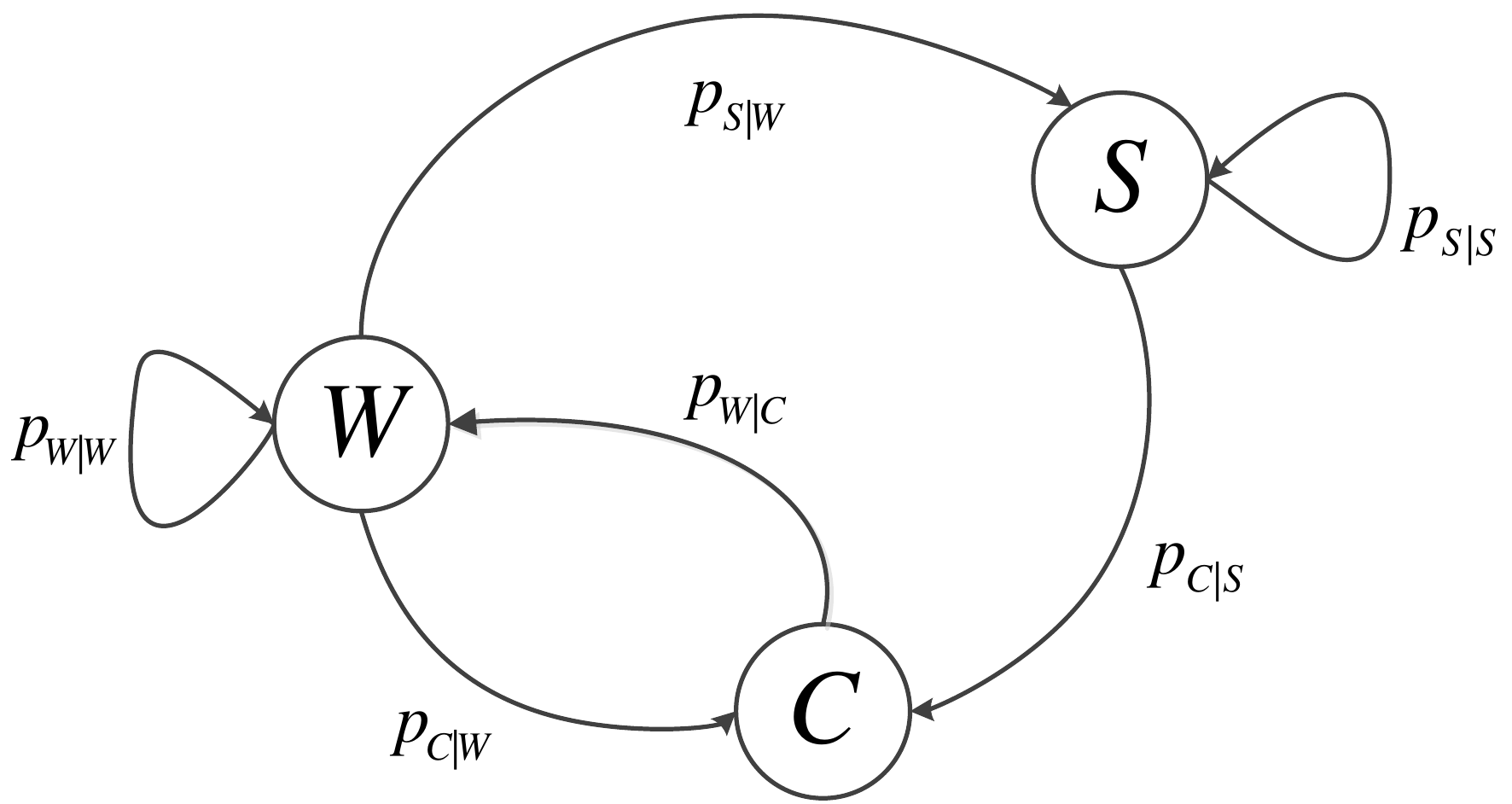}
\includegraphics[height=3.5cm,width=8.8cm]{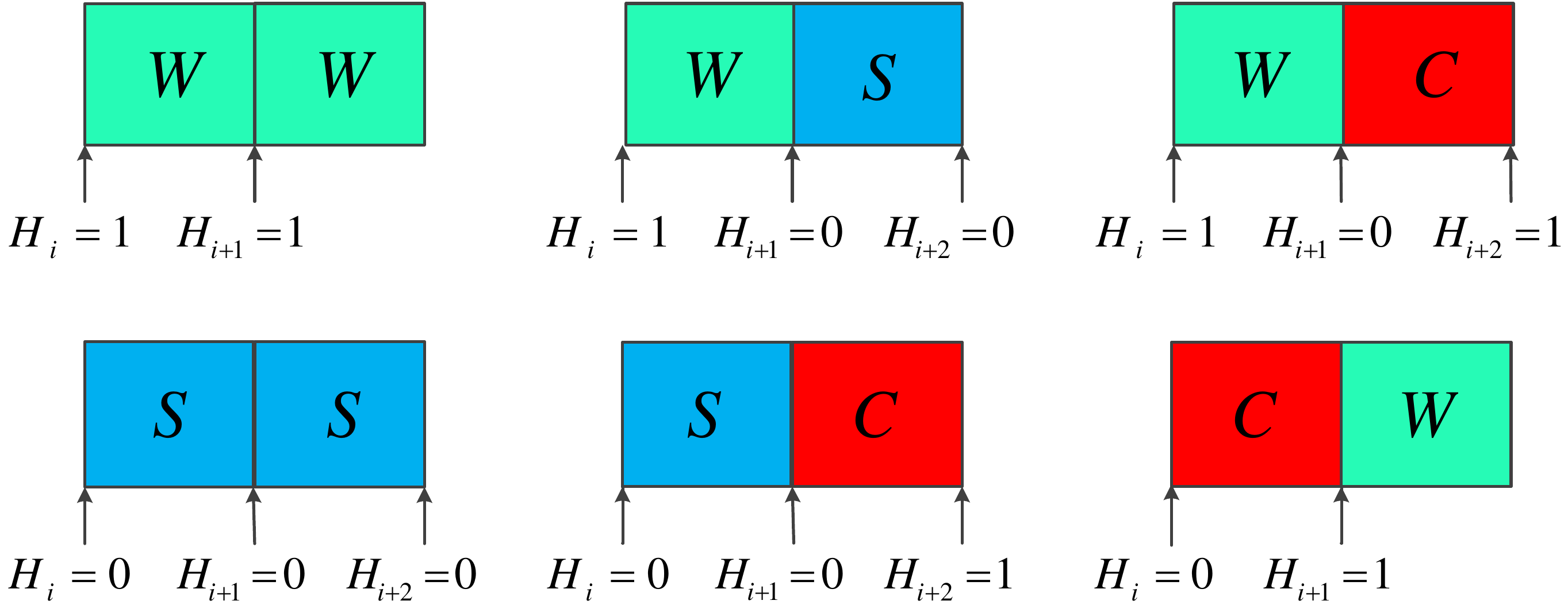}
\caption{Markov modeling of secondary slotted transmission.}
\label{Makov}
\end{center}
\end{figure}

We form a three-state discrete-time Markov chain to characterize SU activity, where state $W$ represents that SU is waiting for transmission over a $T_s$, state $S$ that SU transmits without collision during a $T_s$ and state $C$ that SU transmission collides with PU transmission, as illustrated in Fig. \ref{Makov}. Let $H_i$ denote the sensing result at the start of $i$th $T_s$. Specifically, $H_i=1$ implies that the channel is sensed busy at the start of $i$th $T_s$, and PU is on with the perfect-sensing assumption. $H_i=0$ corresponds that channel is sensed free and PU is off the start of $i$th $T_s$. The transition probability from state $W$ to state $S$ is calculated as
\begin{align}
p_{S|W}&=\Pr[H_{i+1}=0,\ H_{i+2}=0|H_i=1]\notag\\
&=\Pr[H_{i+2}=0|H_{i+1}=0]\Pr[H_{i+1}=0|H_i=1]\notag\\
&=\beta_{off}(1-\beta_{on}),\label{pS|W}
\end{align}
where $\beta_{on}=\Pr[H_i=1|H_{i-1}=1]$ is the probability that PU is sensed on at current sensing instant given PU is also on at previous sensing instant, and $\beta_{off}=\Pr[H_i=0|H_{i-1}=0]$ represents the probability that PU is sensed off given previous sensing result is also off. $\beta_{on}\big/\beta_{off}$ can be calculated as the probability that PU busy/idle duration is larger than $T_s$ and given by \cite{Tscalcula}
\begin{equation}
\beta_{on}=e^{-\frac{T_s}{\lambda}}\ \textrm{and}\  \beta_{off}=e^{-\frac{T_s}{\mu}},\label{beta}
\end{equation}
respectively. Similarly, the transition probability from state $S$ to state $S$ is calculated as
\begin{align}
p_{S|S}
&=\Pr[H_{i+1}=0,\ H_{i+2}=0|H_i=0,\ H_{i+1}=0]\notag\\
&=\Pr[H_{i+2}=0|H_{i+1}=0]\notag\\
&=\beta_{off}.\label{pS|S}
\end{align}
Other transition probabilities can be similarly obtained. Thus, the transition probability matrix of the three-state Markov chain is given by
\begin{align}
\mathbf{P}&=
\begin{bmatrix}
    p_{S|S} & p_{S|W} & p_{S|C} \\
    p_{W|S} & p_{W|W} & p_{W|C} \\
    p_{C|S} & p_{C|W} & p_{C|C}
\end{bmatrix}\notag\\
&=\begin{bmatrix}
    \beta_{off} & (1-\beta_{on})\beta_{off} & 0 \\
    0 & \beta_{on} & 1 \\
    1-\beta_{off} & (1-\beta_{on})(1-\beta_{off}) & 0
\end{bmatrix}.\label{Matrix_3}
\end{align}
\begin{figure}[t]
\begin{center}
\includegraphics[height=7cm,width=8.8cm]{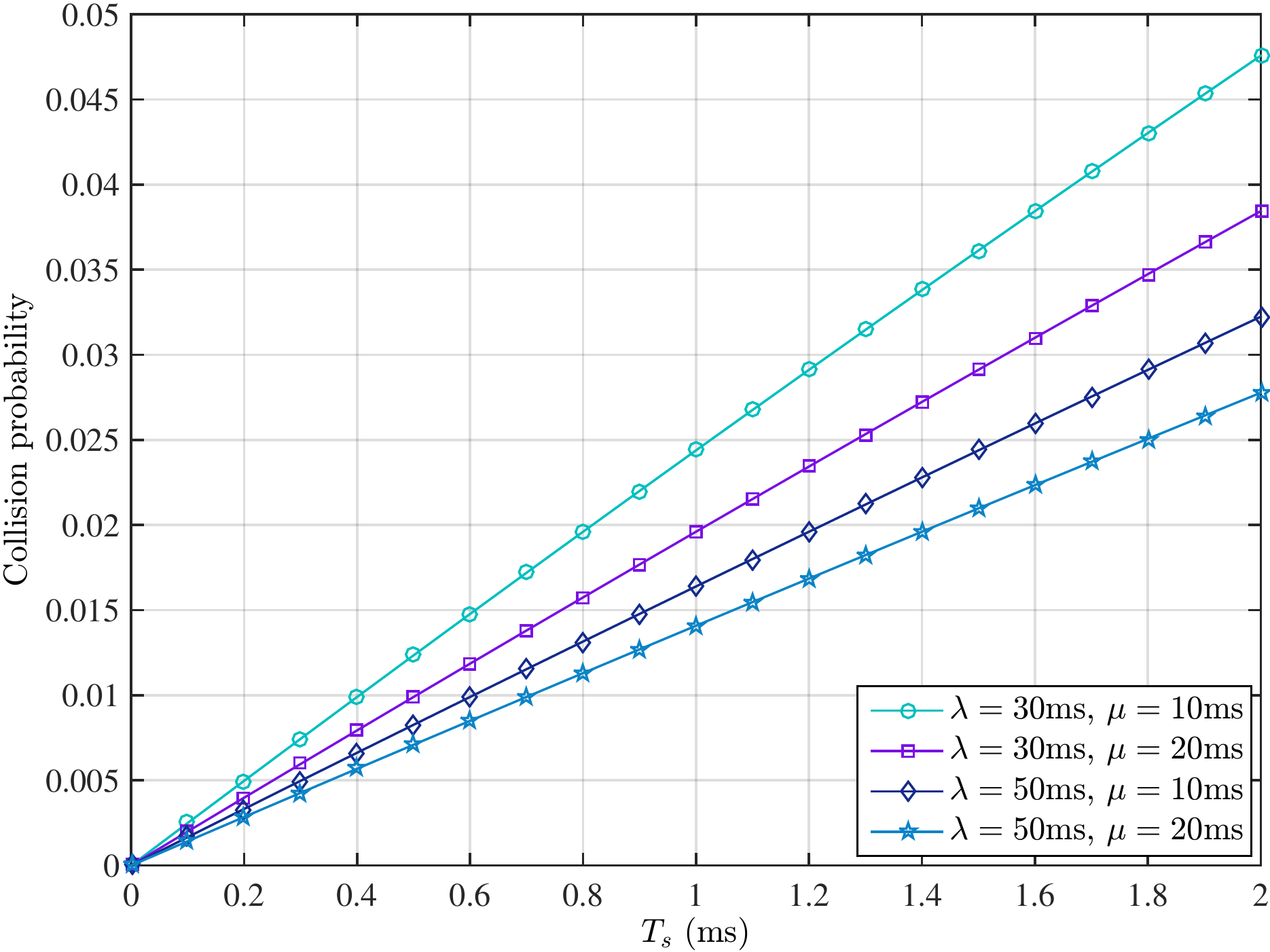}
\end{center}
\caption{Illustration of correspondence of collision probability and sensing period for various PU activities.}\label{CoPr_per}
\end{figure}

\newcounter{TempEqCnt4}
\begin{figure*}[t]
\setcounter{TempEqCnt4}{\value{equation}}
\setcounter{equation}{10}
{\footnotesize
\begin{align}
\mathbf{P}_\textrm{Q}=\left[\begin{array}{ccc;{2pt/2pt}ccc;{2pt/2pt}ccc}
    p_{(S,0)|(S,0)} & \hdots & p_{(S,0)|(S,K)} & p_{(S,0)|(W,0)} & \hdots & p_{(S,0)|(W,K)} & p_{(S,0)|(C,0)} & \hdots & p_{(S,0)|(C,K)}\\
    \vdots & \ddots & \vdots & \vdots & \ddots & \vdots & \vdots & \ddots & \vdots\\
    p_{(S,K)|(S,0)} & \hdots & p_{(S,K)|(S,K)} & p_{(S,K)|(W,0)} & \hdots & p_{(S,K)|(W,K)} & p_{(S,K)|(C,0)} & \hdots & p_{(S,K)|(C,K)}\\
    \hdashline[2pt/2pt]
    p_{(W,0)|(S,0)} & \hdots & p_{(W,0)|(S,K)} & p_{(W,0)|(W,0)} & \hdots & p_{(W,0)|(W,K)} & p_{(W,0)|(C,0)} & \hdots & p_{(W,0)|(C,K)}\\
    \vdots & \ddots & \vdots & \vdots & \ddots & \vdots & \vdots & \ddots & \vdots\\
    p_{(W,K)|(S,0)} & \hdots & p_{(W,K)|(S,K)} & p_{(W,K)|(W,0)} & \hdots & p_{(W,K)|(W,K)} & p_{(W,K)|(C,0)} & \hdots & p_{(W,K)|(C,K)}\\
    \hdashline[2pt/2pt]
    p_{(C,0)|(S,0)} & \hdots & p_{(C,0)|(S,K)} & p_{(C,0)|(W,0)} & \hdots & p_{(C,0)|(W,K)} & p_{(C,0)|(C,0)} & \hdots & p_{(C,0)|(C,K)}\\
    \vdots & \ddots & \vdots & \vdots & \ddots & \vdots & \vdots & \ddots & \vdots\\
    p_{(C,K)|(S,0)} & \hdots & p_{(C,K)|(S,K)} & p_{(C,K)|(W,0)} & \hdots & p_{(C,K)|(W,K)} & p_{(C,K)|(C,0)} & \hdots & p_{(C,K)|(C,K)}\\
\end{array}\right]\label{PQ}
\end{align}
}
\hrulefill
\end{figure*}
\setcounter{equation}{\value{TempEqCnt4}}

The stationary distribution of the Markov chain can be calculated by normalizing the eigenvector of $\mathbf{P}$ corresponding to eigenvalue one as
\begin{align}
&[p_{S},\ p_{W},\ p_{C}]=\notag\\
&\Bigg[\frac{(1-\beta_{on})\beta_{off}}{2-\beta_{on}-\beta_{off}},\ \frac{1-\beta_{off}}{2-\beta_{on}-\beta_{off}},\ \frac{(1-\beta_{on})(1-\beta_{off})}{2-\beta_{on}-\beta_{off}}\Bigg].
\label{Station_3}
\end{align}
%The Markov model can be verified by solving Markov chain at equilibrium. As an illustration, for state $C$
%\begin{equation}
%\begin{aligned}
%&p_{\text{w}}p_{\text{c}|\text{w}}+p_{\text{s}}p_{\text{c}|\text{s}}\\
%=&(1-\beta_{off})(1-\beta_{on})\frac{1-\beta_{off}}{2-\beta_{on}-\beta_{off}}\\
%&\ \ \ \ \ \ \ \ \ \ \ \ \ \ \ +(1-\beta_{off})\frac{(1-\beta_{on})\beta_{off}}{2-\beta_{on}-\beta_{off}}=p_{\text{c}}p_{\text{w}|\text{c}}
%\end{aligned}
%\end{equation}

An immediate application of the Markov chain modeling is to evaluate the collision probability with slotted secondary transmission strategy. Fig. \ref{CoPr_per} plots the collision probability $p_{C}$ as a function of sensing interval $T_s$ for different primary transmission parameters. Larger $\lambda$ or larger $\mu$ results in smaller collision probability. Note that collision only occurs when PU restarts transmission. We also observe that collision probability increases for a longer sensing period as expected by intuition.

\section{Queueing Analysis of Secondary Packet Transmission}
\begin{figure}[t]
\begin{center}
\includegraphics[height=3.1cm,width=8.8cm]{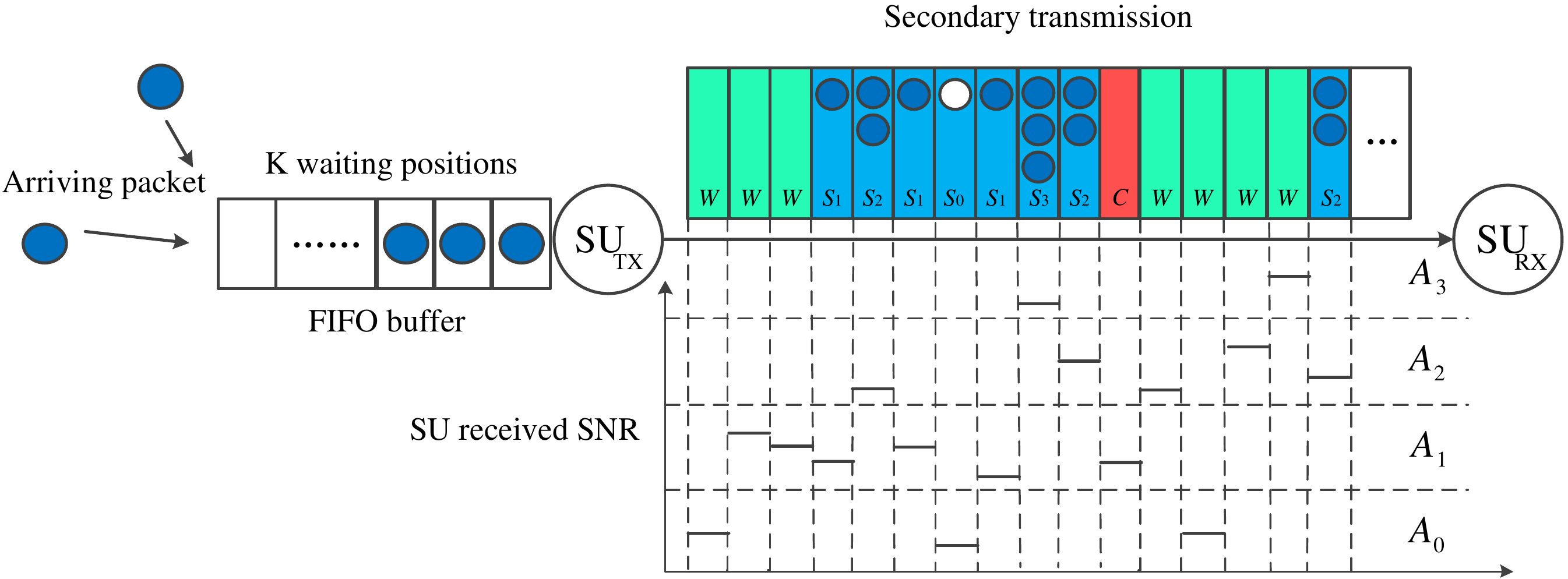}
\end{center}
\caption{Illustration of secondary transmission with AM for small-size packet.}\label{MG1}
\end{figure}
\begin{figure}[b]
\begin{center}
\includegraphics[height=2cm,width=8.8cm]{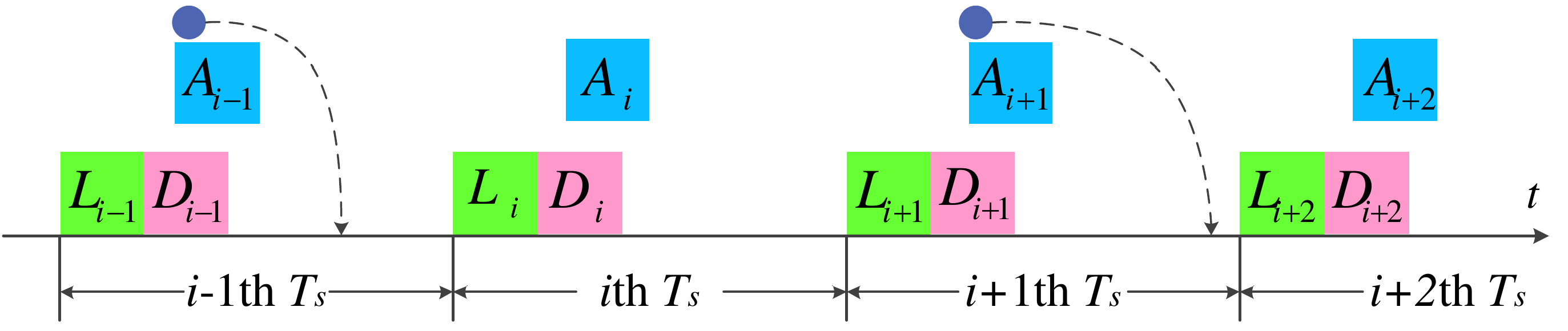}
\end{center}
\caption{Illustration of queue recursion of secondary transmission.}\label{Que_re}
\end{figure}

\begin{figure}[t]
\begin{center}
\includegraphics[height=65mm,width=88mm]{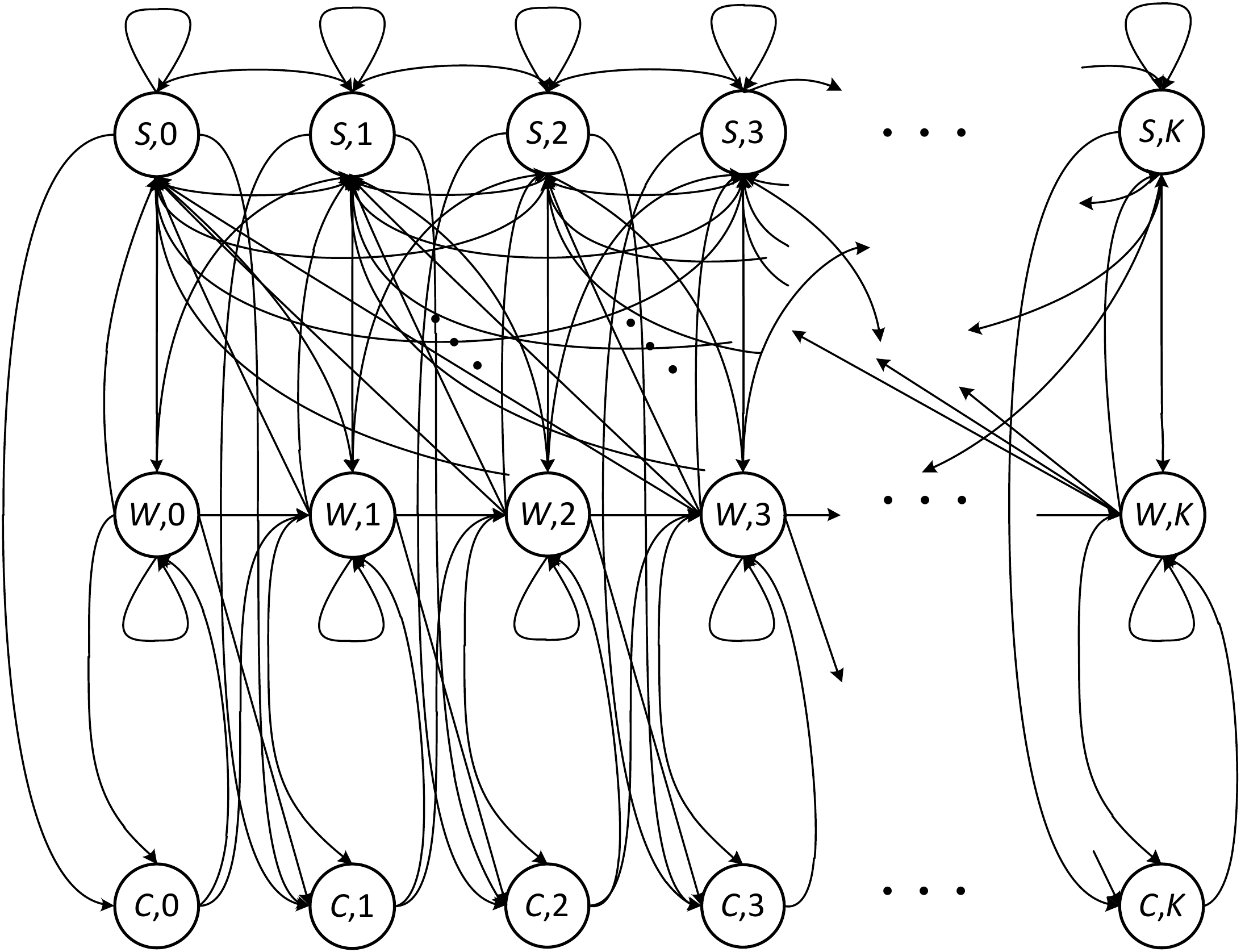}
\end{center}
\caption{Illustration of two-dimensional finite state Markov chain secondary transmission with N-state adaptive modulation.}\label{smalpac}
\end{figure}

\newcounter{TempEqCnt2}
\begin{figure*}[b]
\hrulefill
\setcounter{TempEqCnt2}{\value{equation}}
\setcounter{equation}{14}
{\footnotesize
\begin{equation}
\mathbf{P}_S=
\begin{bmatrix}
    1-p_a & (1-p_a)\sum\limits_{j=1}\limits^{N}\pi_j & (1-p_a)\sum\limits_{j=2}\limits^{N}\pi_j & (1-p_a)\sum\limits_{j=3}\limits^{N}\pi_j & \hdots & 0 \\
    p_a & (1-p_a)\pi_0+p_a\sum\limits_{j=1}\limits^{N}\pi_j & (1-p_a)\pi_1+p_a\sum\limits_{j=2}\limits^{N}\pi_j & (1-p_a)\pi_2+p_a\sum\limits_{j=3}\limits^{N}\pi_j & \hdots & 0\\
    0 & p_a\pi_0 & (1-p_a)\pi_0+p_a\pi_1 & (1-p_a)\pi_1+p_a\pi_2 & \hdots & 0\\
    0 & 0 & p_a\pi_0 & (1-p_a)\pi_0+p_a\pi_1 & \hdots & 0\\
    0 & 0 & 0 & p_a\pi_0 & \hdots  & (1-p_a)\pi_N\\
    \vdots & \vdots & \vdots & \vdots  & \ddots & \vdots\\
    0 & 0 & 0 & 0 & \hdots  & (1-p_a)\pi_1+p_a\pi_2\\ 
    0 & 0 & 0 & 0 & \hdots  & \pi_0+p_a\pi_1 \label{Ps}
\end{bmatrix}
\end{equation}
}
\end{figure*}
\setcounter{equation}{\value{TempEqCnt2}}

In this section, we apply the three-state Markov chain developed in previous section to evaluate the queueing performance of secondary packet transmission. We assume that SU information data is fragmented into equal-size small packets. The packet size is chosen such that $j$ packets can be transmitted over one $T_s$ when the primary channel is available and secondary link can support transmission rate $R_j, j=0,1,\cdots,N$, where $R_0=0$ corresponds to the case that SU decides not to transmit due to unacceptable secondary link quality. The newly arrived packet will be put into a first-in-first-out (FIFO) buffer with $K$ waiting positions before being delivered over the next available transmission slot, as illustrated in Fig. \ref{MG1}. New arrival will be dropped if the queue is full. Only those packets that are successfully received will be removed from the buffer. During each $T_s$, at most one packet may arrive at the SU transmitter with probability $p_a$. Let $\mathcal{A}_{i}\in\{0,1\}$ represent the number of packet arrives during the $i$th $T_s$. We have
\begin{align}
\Pr[\mathcal{A}_i=a]&=\left\{\begin{aligned}&p_a,&a=1;\\
&1-p_a,&a=0.\end{aligned}\right. \label{Pro_arr} 
\end{align}

The service process is independent of the arrival process. We denote the service state of the $i$th $T_s$ as $\mathcal{D}_i$, where $\mathcal{D}_i\in\{C,W,S\}$. When $\mathcal{D}_i=W$, SU waits for channel availability. When $\mathcal{D}_i=C$, collision occurs and no packet was successfully transmitted. When $\mathcal{D}_i=S$ and rate $R_j$ is used, $j$ packets are transmitted successfully. 

Let $\mathcal{L}_i$ denote the instantaneous queue length at the beginning of $i$th $T_s$, where $\mathcal{L}_i\in\{0,1,...,K\}$. As such, $\mathcal{L}_{i}$ depends on $\mathcal{L}_{i-1}$, $\mathcal{D}_{i-1}$ and $\mathcal{A}_{i-1}$, as illustrated in Fig. \ref{Que_re}. Specifically, the queue length will be updated as
\begin{equation}
\mathcal{L}_i=\min\{K,\ \max\{0,\ \mathcal{L}_{i-1}-\mathcal{D}_{i-1}\}+\mathcal{A}_{i-1}\}.\label{Queue_recursion}
\end{equation}

To investigate secondary queuing performance, we construct a two-dimensional Markov chain with state being the $(\mathcal{D}_i,\mathcal{L}_i)$ pair, as illustrated in Fig. \ref{smalpac}. Let $p_{(\mathcal{D}_i,\mathcal{L}_i)|(\mathcal{D}_{i-1},\mathcal{L}_{i-1})}$ denote the transition probability from state $(\mathcal{D}_{i-1},\mathcal{L}_{i-1})$ to state $(\mathcal{D}_i,\mathcal{L}_i)$. Accordingly, the state transition probability matrix, denoted by $\mathbf{P}_\textrm{Q}$, is organized in Eq. (\ref{PQ}), at the top of next page. The transition probabilities can be further simplified, while noting that current service state is only dependent on previous service state, as
\setcounter{equation}{11}
\begin{align}
p_{(\mathcal{D}_i,\mathcal{L}_i)|(\mathcal{D}_{i-1},\mathcal{L}_{i-1})}&=\Pr[\mathcal{D}_i,\mathcal{L}_i|\mathcal{D}_{i-1},\mathcal{L}_{i-1}]\notag\\
%&=\Pr[\mathcal{D}_i|\mathcal{D}_{i-1},\mathcal{L}_{i-1}]\Pr[\mathcal{L}_i|\mathcal{D}_{i-1},\mathcal{L}_{i-1}]\notag\\
&=\Pr[\mathcal{D}_i|\mathcal{D}_{i-1}]\Pr[\mathcal{L}_i|\mathcal{D}_{i-1},\mathcal{L}_{i-1}]. \label{entry_simplified}
\end{align}
$\Pr[\mathcal{D}_i|\mathcal{D}_{i-1}]$ can be calculated by the transition probabilities given in Eq. (\ref{Matrix_3}) in Section III. As such, Eq. (\ref{PQ}) is rewritten in a block form as 
\begin{equation}
\mathbf{P}_\text{Q}=
\begin{bmatrix}
    \beta_{off}\mathbf{P}_S & (1-\beta_{on})\beta_{off}\mathbf{P}_W & \mathbf{O}\\
    \mathbf{O} & \beta_{on}\mathbf{P}_{W} & \mathbf{P}_{C}\\
    (1-\beta_{off})\mathbf{P}_S & (1-\beta_{on})(1-\beta_{off})\mathbf{P}_W & \mathbf{O}
\end{bmatrix},\label{block_pQ}
\end{equation}
where $\mathbf{O}$ represents $(K+1)\times(K+1)$ zero matrix, $\mathbf{P}_S$, $\mathbf{P}_W$ and $\mathbf{P}_C$ are of the form
{\footnotesize
\begin{align}
&\mathbf{P}_{\mathcal{D}_{i-1}}=\notag\\
&\begin{bmatrix}
\Pr[\mathcal{L}_i=0|\mathcal{D}_{i-1},\mathcal{L}_{i-1}=0] & \hdots & \Pr[\mathcal{L}_i=0|\mathcal{D}_{i-1},\mathcal{L}_{i-1}=K]\\
\vdots & \ddots & \vdots\\
\Pr[\mathcal{L}_i=K|\mathcal{D}_{i-1},\mathcal{L}_{i-1}=0] & \hdots & \Pr[\mathcal{L}_i=K|\mathcal{D}_{i-1},\mathcal{L}_{i-1}=K]\\
\end{bmatrix},\label{Ps_definition}
\end{align}}with $\mathcal{D}_{i-1}\in\{C,W,S\}$. As derived in Appendix A, sub-matrices $\mathbf{P}_S$, $\mathbf{P}_W$ and $\mathbf{P}_C$ are given by Eq. (\ref{Ps}) and Eq. (\ref{PwPc}), respectively.
\setcounter{equation}{15}
\begin{align}\notag
&\mathbf{P}_W=\mathbf{P}_C=\\
&\begin{bmatrix}
    1-p_a & 0 & 0 & \hdots & 0 & 0 & 0 & \hdots & 0\\
    p_a & 1-p_a & 0 & \hdots & 0 & 0 & 0 & \hdots & 0\\
    0 & p_a & 1-p_a & \hdots & 0 & 0 & 0 & \hdots & 0\\
    0 & 0 & p_a & \hdots & 0 & 0 & 0 & \hdots & 0\\
    \vdots & \vdots & \vdots & \vdots & \vdots  & \vdots & \vdots & \vdots & \vdots\\
    0 & 0 & 0 & \hdots & 0 & 0 & 0 & \hdots & 1\label{PwPc}
\end{bmatrix},
\end{align}

\begin{figure}
\begin{center}
\includegraphics[height=70mm,width=88mm]{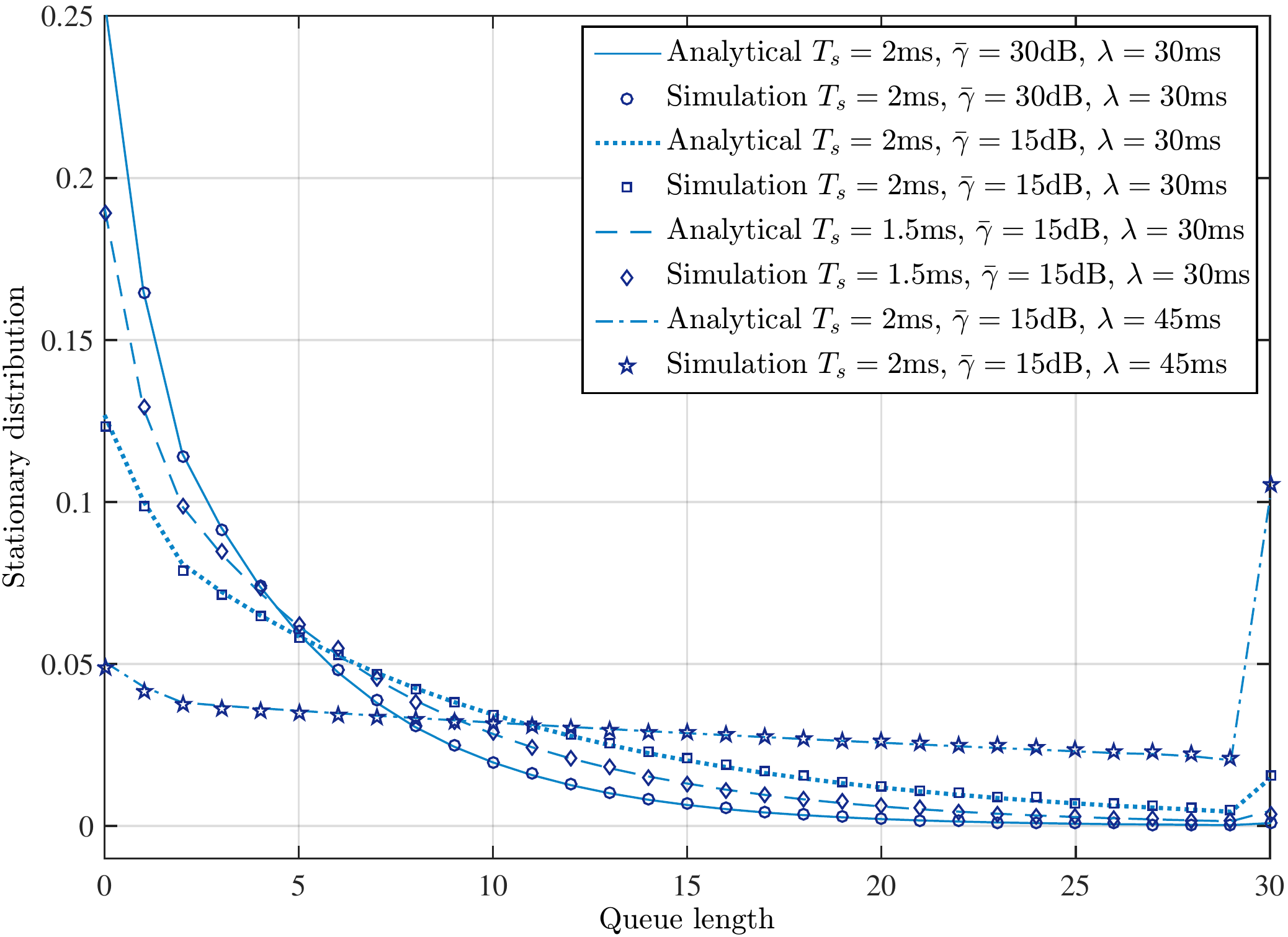}
\end{center}
\caption{Illustration of both analytical and simulation results on the stationary distribution of queue length with different sensing periods, primary user parameters and secondary channel condition, where $K=30$, $\mu=10$ms and $p_a=0.2$.}\label{ql}
\end{figure}

The steady state probabilities of the two-dimensional Markov chain, denoted by
\begin{equation}
\overrightarrow{\varphi_\textrm{Q}}=[\varphi_{S,0},\cdots,\varphi_{S,K},\varphi_{W,0},\cdots,\varphi_{W,K},\varphi_{C,0},\cdots,\varphi_{C,K}],\end{equation}\label{Station_2dimensional}is the left eigenvector of $\mathbf{P}_{\text{Q}}$ corresponding to eigenvalue one, which can be calculated from the forward equations and the normalization equation. The stationary distribution of queue length can be found as $\varphi_k=\sum_{\mathcal{D}=\text{W,C,S}}\varphi_{\mathcal{D},k},\ k=0,1,\cdots,K$.

Fig. \ref{ql} compares the analytical results of stationary distribution of the queue length against corresponding Monte Carlo simulations, where a four-state AM system with transmission rates $[R_0,\ R_1,\ R_2,\ R_3]=[0,\ 1,\ 2,\ 3]$ bits/symbol is used. We also assume that channel gain of secondary link follows Rayleigh distribution. The perfect match here validates our analytical approaches. As average received SNR, denoted by $\bar{\gamma}$, gets larger, higher-order modulation schemes are used more frequently. Hence, $\varphi_k$ is larger for smaller $k$. For larger $T_s$, SU transmits less packets while PU is idle, which results in higher $\varphi_k$ for larger $k$. $\varphi_k$ for larger $k$ decreases with decreasing $\lambda$ since SU has more chances to acquire the channel. 
\begin{figure}
\begin{center}
\includegraphics[height=70mm,width=88mm]{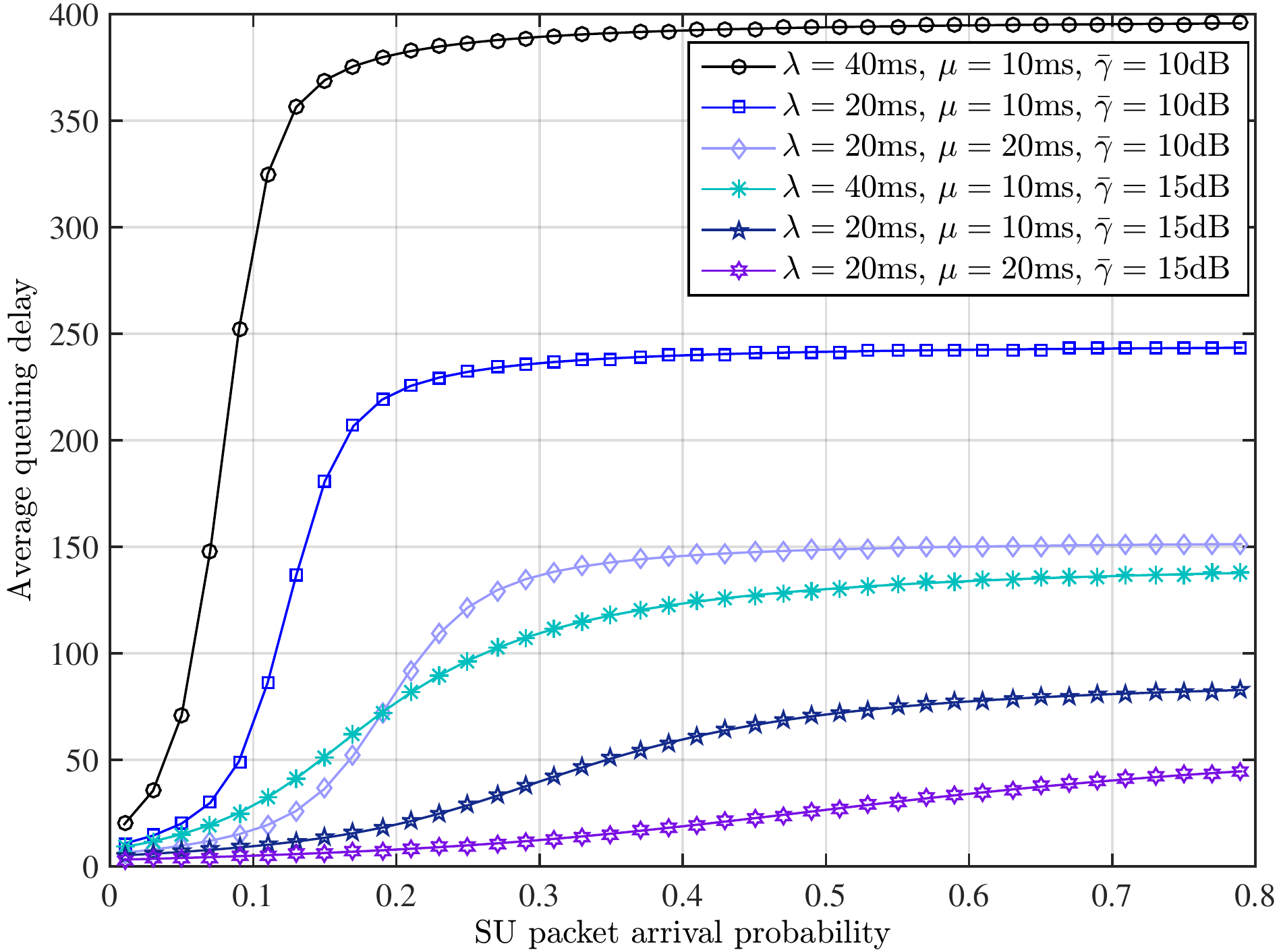}
\end{center}
\caption{Illustration of the average queuing delay versus various secondary channel quality and primary user parameters.}\label{Que_qd}
\end{figure}
\begin{figure}
\begin{center}
\includegraphics[height=70mm,width=88mm]{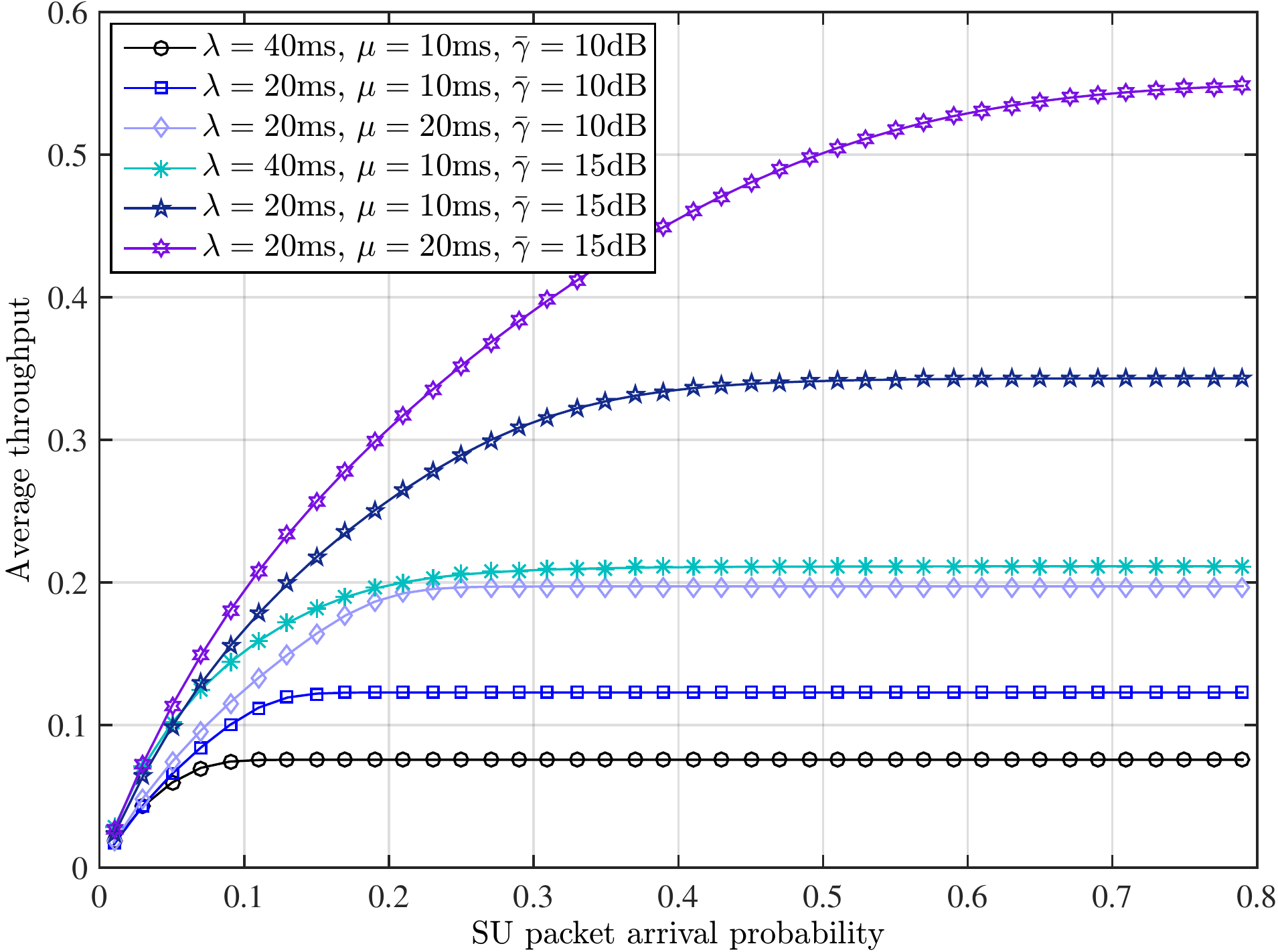}
\end{center}
\caption{Illustration of the average throughput versus various secondary channel quality and primary user parameters.}\label{Que_thr}
\end{figure}
\begin{figure}
\begin{center}
\includegraphics[height=70mm,width=88mm]{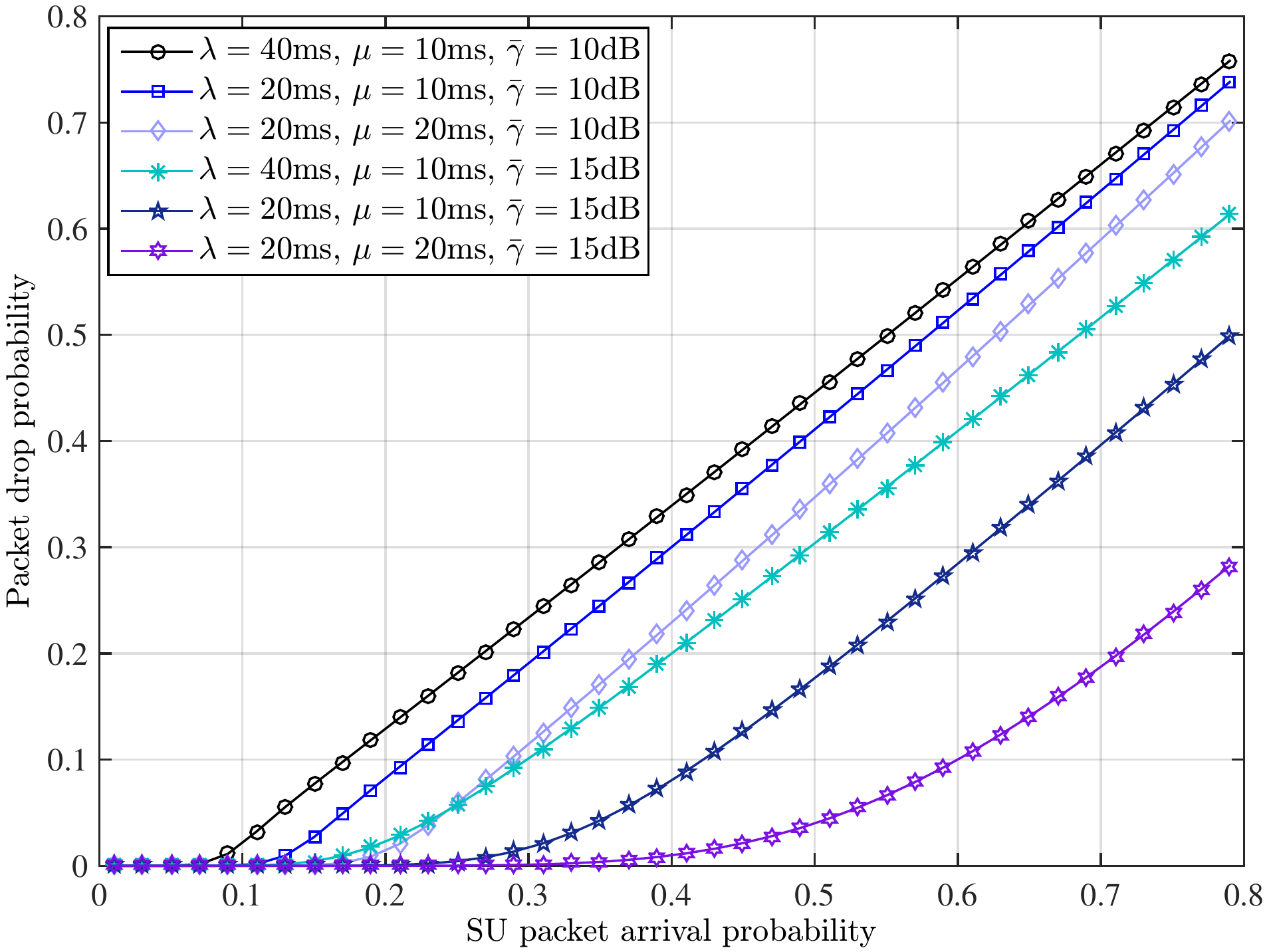}
\end{center}
\caption{Illustration of the packet drop probability versus various secondary channel quality and primary user parameters.}\label{Que_pdp}
\end{figure}

We evaluate the queuing performance in term of packet drop probability, throughput, and queuing delay. Packet drop occurs when arriving packet finds that the queue is full, the probability of which is calculated as
\begin{equation}
P_{\text{drop}}=\varphi_Kp_a. \label{Packet_drop}
\end{equation} 
The average throughput of secondary system, denoted by $\mathbb{E}[Th]$, can be calculated as
\begin{equation}
\mathbb{E}[Th]=p_S\sum_{k=0}^K\mathbb{E}[\mathcal{S}|k]\varphi_k, \label{Throughput}
\end{equation}
where $p_S$ is the stationary probability that SU transmits without collision, given by Eq. (\ref{Station_3}), and $\mathbb{E}[\mathcal{S}|k]$ is the average number of packets been successfully transmitted per $T_s$ given $k$ packets waiting in the queue, which can be calculated as
\begin{align}
\label{Conditional_deliver}
\mathbb{E}[\mathcal{S}|k]=
\left\{\begin{aligned}
&0,&k=0;\\
&\sum_{i=1}^ki\pi_i+k\sum_{j=k+1}^N\pi_j, &0< k< N;\\
&\sum_{i=1}^Ni\pi_i,&N\leq k\leq K.\\
\end{aligned}\right.
\end{align}
The average queue-length, denoted by $\mathbb{E}[Q]$, can be calculated as $\mathbb{E}[Q]=\sum_{k=0}^Kk\varphi_k$. Applying Little's law, the average queuing delay can be calculated as
\begin{equation}
\mathbb{E}[T_Q]=\frac{\mathbb{E}[Q]}{\mathbb{E}[Th]}=\frac{\sum_{k=0}^Kk\varphi_k}{p_S\sum_{k=0}^K\mathbb{E}[\mathcal{S}|k]\varphi_k}.\label{Queuing_delay}
\end{equation}

In Fig. \ref{Que_qd}, we plot the average queuing delay as a function of arriving probability. The average delay saturates as input traffic intensity grows because packets are dropped when queue is full. For smaller $\mu$ or larger $\lambda$, SU has less chances to employ the channel leading to a larger saturate value. As $\bar{\gamma}$ increases, SU transmits with higher rate more frequently, which helps reduce the queuing delay. Fig. \ref{Que_thr} shows the average throughput versus arriving probability for various PU parameter and secondary link quality. For larger $\mu$, smaller $\lambda$ or lager $\bar{\gamma}$, SU can transmit for longer period or more packets on average. As such, average throughput saturates to a larger value at a lager arriving probability. When average throughput approaches its maximum, packet drop probability increases sharply as expected by intuition (see Fig. \ref{Que_pdp}).

\newcounter{TempEqCnt3}
\begin{figure*}[!hb]
\hrulefill
\setcounter{TempEqCnt3}{\value{equation}}
\setcounter{equation}{23}
{\footnotesize
\begin{align}
\Pr[T_{\textrm{ED}}=LT_s]=&\frac{(1-\beta_{on})\beta_{off}}{2-\beta_{on}-\beta_{off}}\Bigg\{\sum_{i=1}^{\phi-1}\sum_{\begin{subarray}{c}m,\ j\ \text{s.t.}\\m+2j=L-\phi-2i\end{subarray}}\Bigg[\binom{m+i-1}{i-1,\ j}p_{S|W}^ip_{C|W}^jp_{W|W}^{m-j}\binom{\phi-1}{i}p_{S|S}^{\phi-1-i}p_{C|S}^{i}\Bigg]+\delta(\phi)p_{S|S}^{\phi-1}\Bigg\}\notag\\
+&\frac{1-\beta_{off}}{2-\beta_{on}-\beta_{off}}\sum_{i=1}^{\phi}\sum_{\begin{subarray}{c}m,\ j\ \text{s.t.}\\m+2j=L-1-\phi-2(i-1)\end{subarray}}\binom{m+i-1}{i-1,j}p_{S|W}^ip_{C|W}^jp_{W|W}^{m-j}\binom{\phi-1}{i-1}p_{S|S}^{\phi-i}p_{C|S}^{i-1}\notag\\
+&\frac{(1-\beta_{on})(1-\beta_{off})}{2-\beta_{on}-\beta_{off}}\sum_{i=1}^{\phi}\sum_{\begin{subarray}{c}m,\ j\ \text{s.t.}\\m+2j=L-2-\phi-2(i-1)\end{subarray}}\binom{m+i-1}{i-1,j}p_{S|W}^ip_{C|W}^jp_{W|W}^{m-j}\binom{\phi-1}{i-1}p_{S|S}^{\phi-i}p_{C|S}^{i-1}
\label{PMF_Ted_PS}
\end{align}
}
\end{figure*}
\setcounter{equation}{\value{TempEqCnt3}}

\section{EDT analysis for Large-size packet transmission}
For large-size packet, multiple $T_s$'s are required to complete single packet transmission. As such, packet service time consists of waiting time and transmission time. We derive the statistics of EDT assuming fixed-rate transmission and variable-rate transmission in the following subsections.
\subsection{Fixed-Rate Transmission}
\begin{figure}
\begin{center}
\includegraphics[height=7cm,width=8.8cm]{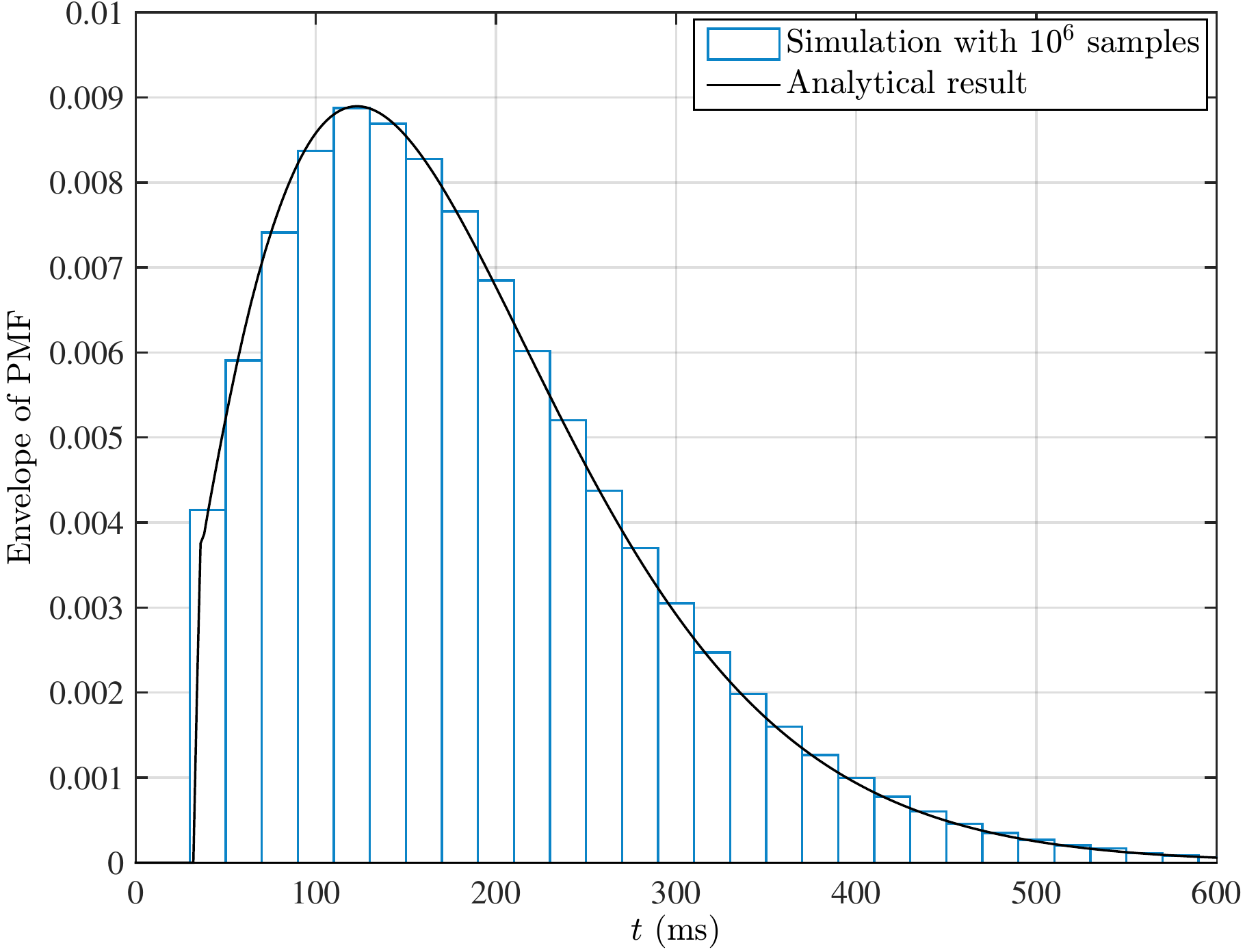}
\end{center}
\caption{Monte Carlo verification of $T_{\textrm{ED}}$ analysis ($H_t=8$kb, $\lambda=30$ms, $\mu=10$ms, $\bar{\gamma}=20$dB).}\label{PDFQPSK}
\end{figure}

\begin{figure}
\begin{center}
\includegraphics[height=7cm,width=8.8cm]{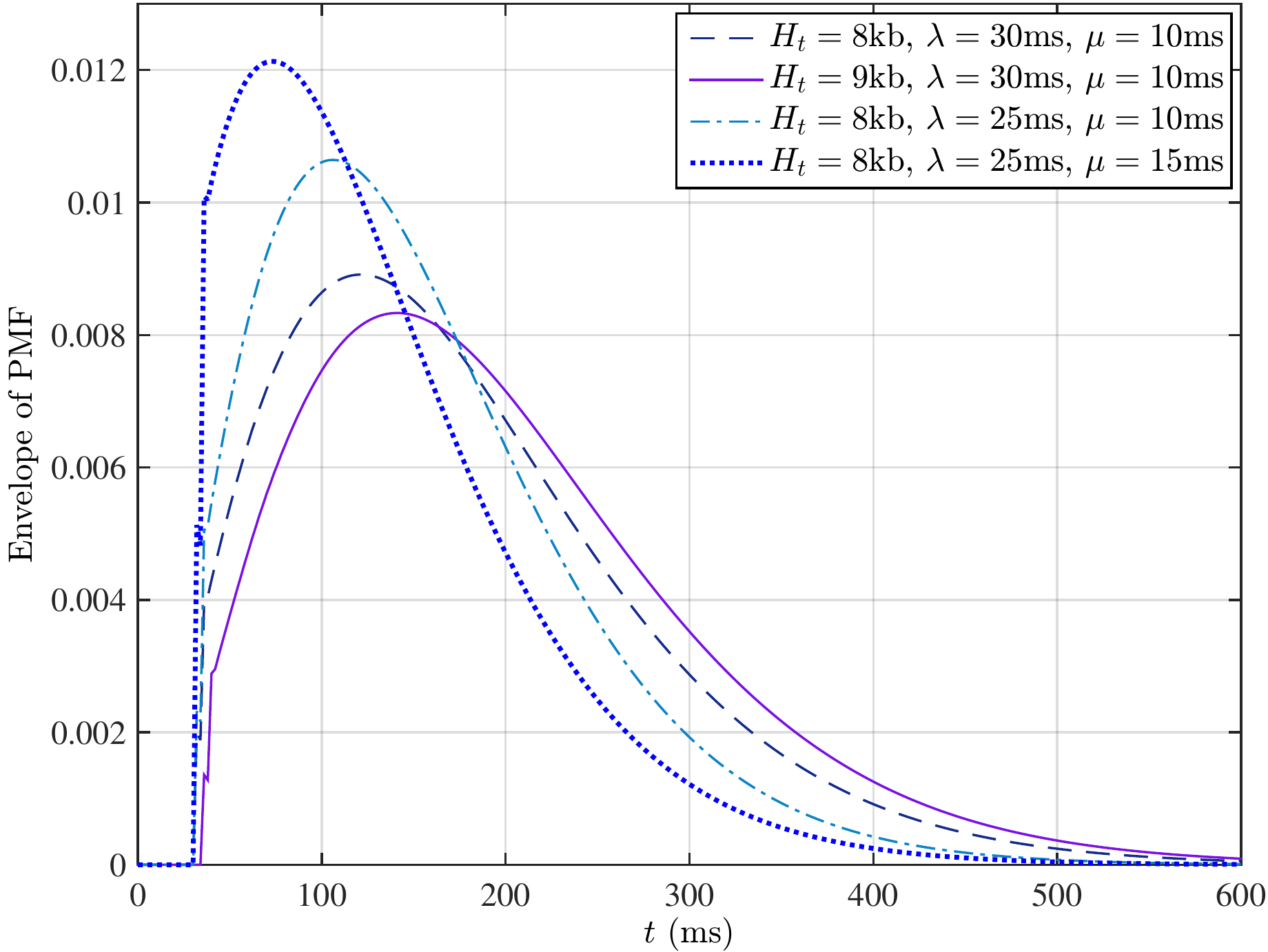}
\end{center}
\caption{Illustration of  PMF of EDT with fixed-rate transmission for various PU activities and secondary packet size.}\label{ComTed}
\end{figure}

Firstly, we derive the PMF of EDT ($T_{\textrm{ED}}$) for a fixed-size packet transmission with a fixed data rate. $T_{\textrm{ED}}$ should include all the waiting intervals, collision intervals and transmission intervals before transmission completed. Suppose that transmission is accomplished over $\phi$ transmission intervals, which can be approximated by
\begin{equation}
\phi=\Bigg\lceil\frac{H_t}{RT_s}\Bigg\rceil, \label{Srequired}
\end{equation}
where $H_t$ represents the entropy of the packet, and $R$ is the data rate. Specifically, the PMF of $T_{\textrm{ED}}$ depends on the state of the first $T_s$ at the instant of packet arrival. Conditioning on that the first $T_s$ is in state $S$, $W$ and $C$. The PMF of $T_{\textrm{ED}}$ can be calculated as
\begin{align}
\Pr[T_{\textrm{ED}}&=LT_s]=\notag\\
&p_S\Pr\left[T_{\textrm{ED}|S}=LT_s\right]+p_W\Pr\left[T_{\textrm{ED}|W}=LT_s\right]\notag\\
+&p_C\Pr\left[T_{\textrm{ED}|C}=LT_s\right],\label{Ted_formulation_PS}
\end{align}
where $\Pr[T_{\textrm{ED}|S}=LT_s]$, $\Pr[T_{\textrm{ED}|W}=LT_s]$ and $\Pr[T_{\textrm{ED}|C}=LT_s]$ denote the conditional PMF of $T_{\textrm{ED}}$ given that the first $T_s$ is in state $S$, $W$ and $C$, respectively, which are derived in Appendix B. The PMF of $T_{\textrm{ED}}$ for secondary slotted transmission with fixed data rate can be calculated by Eq. (\ref{PMF_Ted_PS}) revealed at the bottom of nest page. Fig. \ref{PDFQPSK} compares the analytical result of $T_{\textrm{ED}}$ and corresponding Monte Carlo simulation. The plot shows that the analytical result conform to the simulation result. Fig. \ref{ComTed} shows the PMF envelope of EDT with periodic sensing for various values of packet size and PU parameters. Both average value and variance of $T_{\textrm{ED}}$ increases with $H_t$. Larger size packet requires more transmission intervals (i.e. a larger $\phi$ is estimated by Eq. (\ref{Srequired})), which leads to even more waiting intervals and collision intervals before transmission completed. When SU has more chance to acquire the channel because of the larger $\mu$ or smaller $\lambda$, both average and variance of $T_{\textrm{ED}}$ decreases. This is mainly because, in a $T_{\textrm{ED}}$ sequence, the number of waiting intervals reduces with increasing $\mu$ or decreasing $\lambda$. 

\subsection{Variable-Rate Transmission}
When AM is used, transmission intervals required to complete transmission is no longer constant but a random variable depending on secondary channel realization. $\phi$ transmission intervals are needed only if data transmitted over $\phi-1$ transmission intervals is less than packet size $H_t$, whereas data transmitted over $\phi$ transmission intervals is greater than the packet size. The probability that transmission time, denoted by $T_{\textrm{tr}}$, is equal to $\phi$ $T_s$'s can be formulated as
\setcounter{equation}{24}
\begin{equation}
\Pr[T_{\textrm{tr}}=\phi T_s]=\Pr\left[H_t-R^{(\phi)}T_s\leq \mathbb{H}_{\phi-1}\leq H_t\right],\label{Ttr_PS}
\end{equation}
where $R^{(\phi)}$ denotes the data rate used in the last transmission interval. As such, $R^{(\phi)}$ is a random variable (i.e. $R^{(\phi)}=R_j$ with probability $\pi_j,\ j=0,1,\cdots,N$). $\mathbb{H}_{\phi-1}$ denotes the data transmitted over the first $\phi-1$ transmission intervals. If secondary user received SNR falls into region $A_l$ at the $\phi$th interval, $R_l$ is used (i.e. $R^{(\phi)}=R_l$). By conditioning on the channel realization of the $\phi$th transmission interval, we have
\begin{equation}
\Pr\left[T_{tr}=\phi T_s\right]=\sum_{l=0}^N\Pr\left[H_t-R_lT_s\leq \mathbb{H}_{\phi-1}\leq H_t\right]\pi_l. \label{Condition_on_last_PS}
\end{equation}
Suppose that, during the first $\phi-1$ transmission intervals, SU received SNR falls into region $A_j$ in total $n_j$ times, $j=0,1,\cdots,N$, where $n_j$'s satisfy $\sum_{j=0}^Nn_j=\phi-1$. Let $\overrightarrow{n}=[n_0,\ n_1,\cdots,\ n_N]$ represent one certain channel realization over the first $\phi-1$ transmission intervals. Applying the result of multinomial distribution, we arrive at
\begin{equation}
\Pr\left[\mathbb{H}_{\phi-1}=T_s\sum_{j=0}^Nn_jR_j\right]=\binom{\phi-1}{n_0,n_1,...,n_N}\prod_{j=0}^N\pi_j^{n_j}.\label{Pro_Hk-1}
\end{equation}
Note the independence between different channel realizations, we have
\begin{equation}
\begin{aligned}
\Pr\big[H_t-R_l&T_s\leq \mathbb{H}_{\phi-1}\leq H_t\big]\\
=&\sum_{\overrightarrow{n}}\Pr\left[H_t-R_lT_s\leq \mathbb{H}_{\phi-1}\leq H_t\right]\\
=&\sum_{\begin{subarray}{c}
\overrightarrow{n}\ \text{s.t.}\\
\mathbb{H}_{\phi-1}\in [H_t-R_lT_s,\ H_t]\\
\end{subarray}}\binom{\phi-1}{n_0,n_1,...,n_N}\prod_{j=0}^N\pi_j^{n_j}.\label{Pro_L-1}
\end{aligned}
\end{equation}
The inner sum in Eq. (\ref{Pro_L-1}) is carried out over all $\overrightarrow{n}$'s satisfying that $\mathbb{H}_{\phi-1}$ falls into region $[H_t-R_lT_s,\ H_t]$. Substituting Eq. (\ref{Pro_L-1}) into Eq. (\ref{Condition_on_last_PS}), the probability of $\phi$ transmission slots required is given by \cite{wang_ttr}
\begin{align}
&\Pr[T_{\textrm{tr}}=\phi T_s]=\notag\\
&\sum_{l=0}^N\left(\sum_{\begin{subarray}{c}
\overrightarrow{n}\ \text{s.t.}\\
\mathbb{H}_{\phi-1}\in [H_t-R_jT_s,\ H_t]\\
\end{subarray}}\binom{\phi-1}{n_0,n_1,...,n_N}\prod_{j=0}^N\pi_j^{n_j}\right)\pi_l.
\label{Ttr_Dis_SP}
\end{align}
\begin{figure}[t]
\begin{center}
\includegraphics[height=70mm,width=88mm]{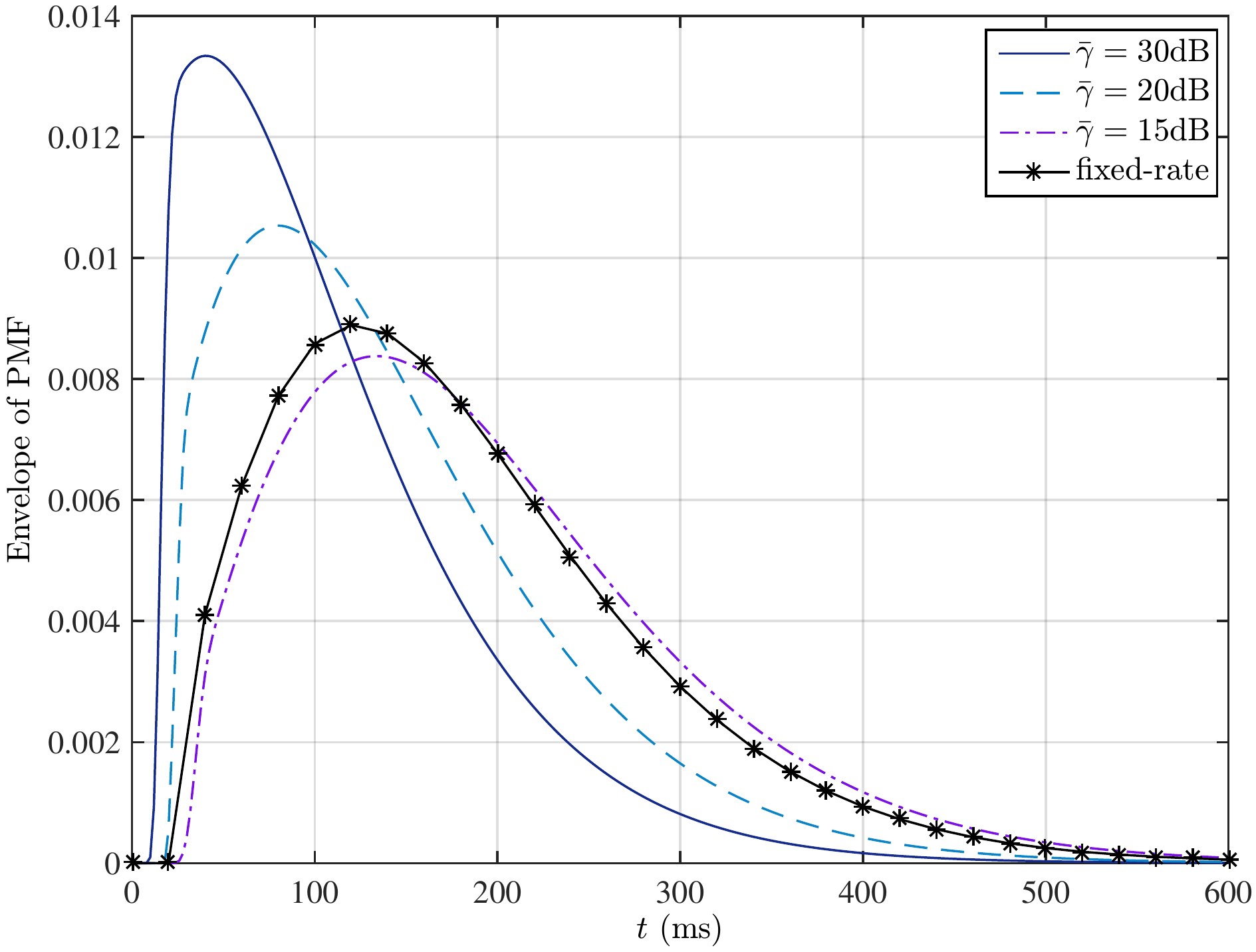}
\end{center}
\caption{Illustration of  PMF of EDT for secondary transmission with four-state AM.}\label{PMFadap}
\end{figure} 

The distribution of $T_{\textrm{ED}}$ can be obtained by conditioning on $\phi$ as
\begin{equation}
\Pr[T_{\textrm{ED}}=LT_s]=\sum_{\phi=1}^{\infty}\Pr[T_{\textrm{ED}|\phi}=LT_s]\Pr[T_{tr}=\phi T_s],\label{Ted_AM_PS}
\end{equation}
where $\Pr[T_{\textrm{ED}|\phi}=LT_s]$ is the distribution of EDT given $\phi$ intervals required, shown in Eq. (\ref{PMF_Ted_PS}). In practical, the value of $\phi$ can not be infinity, since the packet size $H_t$ is fixed. Fig. \ref{PMFadap} compares the envelope of PMF of $T_{\textrm{ED}}$ versus various secondary channel quality. As channel quality gets better, both average and variance of $T_{\textrm{ED}}$ decreases since less intervals are required to complete transmission. Note that, when channel quality is poor, more time is required to accomplish transmission even with AM, compared to fixed-rate transmission. This is because that lower rates are frequently used, or SU decides not to transmit mostly.

\section{Conclusion}
In this paper, we studied packet transmission performance of slotted secondary transmission. We firstly proposed a three-state discrete-time Markov chain to model secondary slotted transmission with periodic spectrum sensing. The stationary probabilities that SU waits/transmits/collides with PU transmission are calculated. The Markov model is then applied to analyze the delay performance of secondary packet transmission. For short packet transmission, a two dimensional Markov chain is formed to evaluate the queuing delay, average throughput and packet drop probability assuming SU adopts adaptive modulation. When packet size is large, we derive the PMF of EDT considering both fixed-rate and variable-rate transmission, respectively. Ongoing effort is being carried out to multiple-PU-multiple-SU cognitive radio network and imperfect sensing scenarios.
\appendix
\section{Derivation of transition probability for two-dimensional Markov chain}
In this appendix, we derive the transition probabilities in $\mathbf{P}_S$, $\mathbf{P}_W$ and $\mathbf{P}_C$, respectively. We firstly derive each entry in matrix $\mathbf{P}_S$, denoted by $\Pr[\mathcal{L}_i=m|S,\mathcal{L}_{i-1}=n]$. Note that $j$ packets are successfully transmitted when rate $R_j$ is used, which is selected with probability $\pi_j,\ j=0,1,\cdots,N$. For $m=0$, given $\mathcal{D}_{i-1}=S$ and $\mathcal{L}_{i-1}=n,\ n=0,1,\cdots,K$, the queue can not be emptied even the highest rate is used if $n>N$. If $n\leq N$, $\mathcal{L}_i=0$ only when the secondary link can support rate $n$ or higher as well as no packet arrives in the $i-1$th $T_s$. Hence, the entries in first row in $\mathbf{P}_S$ can be calculated by
\begin{align}
\Pr[\mathcal{L}_i=0|S,\mathcal{L}_{i-1}&=n]=\left\{\begin{aligned}&0,&n>N;\\
(1-p_a&)\sum_{j=n}^{N-1}\pi_j, &n\leq N.\end{aligned}\right. 
\end{align}\label{Row1} 

If $n=0$, based on the queue recursion, $m=1$ when one packet arrives in the $i-1$th $T_s$. If $1\leq n\leq N$, $m=1$ only when $n-1$ departures and no arrival, or at least $n$ packets departed and one packet arrives. If $n=N+1$, $m=1$ only when highest rate, $R_N$, is used and no packet arrives. For $n>N+1$, $m$ can not be one. Then, the entries in second row can be calculated as 
\begin{align}
\Pr[&\mathcal{L}_i=1|S,\mathcal{L}_{i-1}=n]=\notag\\
&\left\{\begin{aligned}&p_a,&n=0;\\
&(1-p_a)\pi_{n-1}+p_a\sum_{j=n}^{N-1}\pi_j, 1\leq &n\leq N;\\
&(1-p_a)\pi_N, &n=N+1;\\
&0,&n>N+1.\end{aligned}\right.
\end{align}\label{Row2}

Following similar analytical step, when $2\leq m\leq K$, the entries can be calculated by
\begin{align}
&\Pr[\mathcal{L}_i=m|S,\mathcal{L}_{i-1}=n]=\notag\\
&\left\{\begin{aligned}&0,&n-m>N-1;\\
& &\text{or}\ n-m<-1;\\
&(1-p_a)\pi_{n-m},&n-m=N-1;\\
&(1-p_a)\pi_{n-m}+p_a\pi_{n-(m-1)},&0\leq n-m<N-1;\\
&p_a\pi_0, &n-m=-1.\end{aligned}\right.
\end{align}\label{General}
The special case $m=n=K$ can be calculated by
\begin{equation}
\Pr[\mathcal{L}_i=K|S,\mathcal{L}_{i-1}=K]=\pi_0+p_a\pi_1.
\end{equation}
Substituting all the entries in Eq. (\ref{Ps_definition}), $\mathbf{P}_S$ is given by Eq. (\ref{Ps}).

If $\mathcal{D}_{i-1}=W$ or $\mathcal{D}_{i-1}=C$, the queue length would remain unchanged or add up one, depending absolutely on the arrival process. Hence, 
\begin{align}
\Pr[\mathcal{L}_i=m|W,\ &\mathcal{L}_{i-1}=n]=\Pr[\mathcal{L}_i=m|C,\ \mathcal{L}_{i-1}=n]=\notag\\
&\left\{\begin{aligned}
&p_{a},&0\leq m=n+1\leq K;\\
&1-p_{a},&0\leq m=n< K;\\
&1,&m=n=K;\\
&0,&\text{otherwise}.
\end{aligned}\right.
\end{align}\label{31}
As such, the corresponding $(K+1)\times (K+1)$ transition matrix given SU waits or collides with PU transmission at current $T_s$ is given by Eq. (\ref{PwPc}).

\section{Derivation of PMF of EDT for large packet transmission}
\begin{figure}[t]
\begin{center}
\includegraphics[height=5.75cm,width=8.8cm]{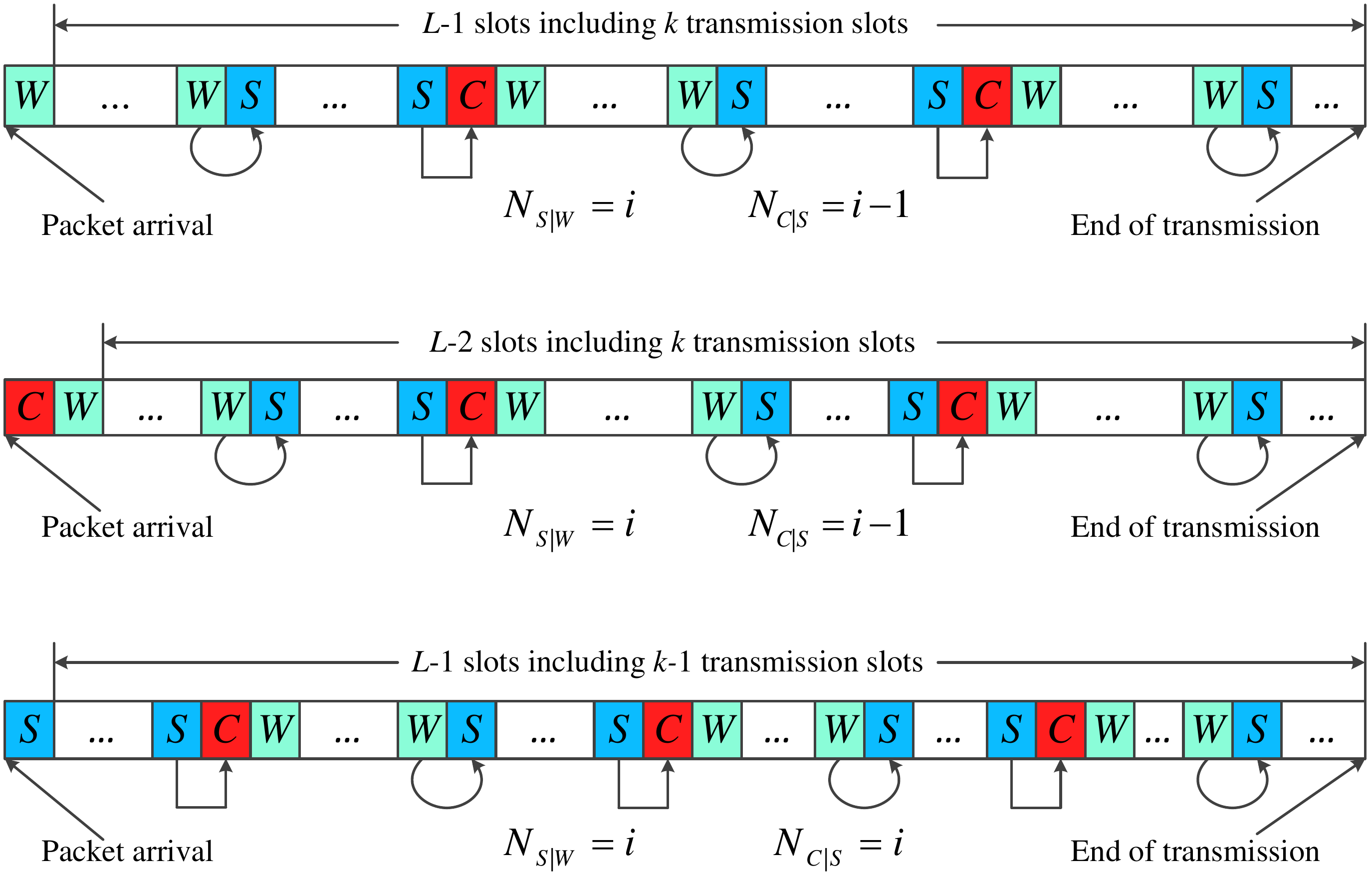}
\end{center}
\caption{Illustration of EDT sequence conditioning on the first slot is in state $W$, $C$ and $S$, respectively.}\label{Ted_Con}
\end{figure}

In this appendix, we derive the conditional PMF of EDT of packet transmission given the first interval is in state $W$, $C$ and $S$ respectively, as illustrated in Fig. \ref{Ted_Con}. Let $N_{\mathcal{D}_i|\mathcal{D}_{i-1}}$ denote the amount of $\mathcal{D}_{i}$ intervals given that the previous interval is in state $\mathcal{D}_{i-1}$. Note that, once collision occurs, the following interval would be a waiting interval, we have $N_{W|C}=N_{C|W}+N_{C|S}$. And when the number of transmission intervals is estimated by Eq. (\ref{Srequired}), $N_{S|S}+N_{S|W}=\phi$.
Suppose that, during SU packet transmission, PU switches from busy to idle $i$ times in total, $i=1,2,\cdots,\phi$. Conditioning on starting with a $W$ slot, we have $N_{S|W}=i$, $N_{S|S}=\phi-i$ and $N_{C|S}=i-1$. And $\Pr[T_{\textrm{ED}|W}=LT_s]=\Pr[N_{S|S}+N_{C|S}+N_{S|W}+N_{W|W}+N_{C|W}+N_{W|C}=L-1]$. The conditional PMF of $T_{\textrm{ED}|W}$ can be further simplified as
\begin{align}
&\Pr[T_{\textrm{ED}|W}=LT_s]\notag\\
=&\Pr[\phi+N_{W|W}+2N_{C|W}+2N_{C|S}=L-1]\notag\\
=&\sum_{i=1}^{\phi}\Pr[\phi+N_{W|W}+2N_{C|W}+2(i-1)=L-1]\notag\\
&\ \ \times\Pr[N_{S|W}=i].
\label{33}
\end{align}
Applying binomial distribution, we have
\begin{equation}
\Pr[N_{S|W}=i]=\Pr[N_{C|S}=i-1]=\binom{\phi-1}{i-1}p_{S|S}^{\phi-i}p_{C|S}^{i-1}.\label{34}
\end{equation}
Conditioning on $N_{S|W}=i$, it takes $m+i$ sensing instants to find out secondary transmission can start or restart $i$ times. Suppose $N_{C|W}=j,\ 1\leq j\leq m-i$, applying negative-multinomial distribution, we have
\begin{align}
&\Pr[N_{W|W}+2N_{C|W}=L-1-\phi-2(i-1)]=\notag\\
&\sum_{\begin{subarray}{c}m,\ j\ \text{s.t.}\\m+2j=L-1-\phi-2(i-1)\end{subarray}}\binom{m+i-1}{i-1,\ j}p_{S|W}^ip_{C|W}^jp_{W|W}^{m-j}.
\label{35}
\end{align}
Substituting Eq. (\ref{34}) and Eq. (\ref{35}) into Eq. (\ref{33}), we have
\begin{align}
&\Pr[T_{\textrm{ED}|W}=LT_s]=\notag\\
&\sum_{i=1}^{\phi}\sum_{\begin{subarray}{c}m,\ j\ \text{s.t.}\\m+2j=L-1-\phi-2(i-1)\end{subarray}}\Bigg[\binom{m+i-1}{i-1,\ j}p_{S|W}^ip_{C|W}^jp_{W|W}^{m-j}\notag\\
&\ \ \ \ \ \ \ \ \ \ \ \ \ \ \ \ \ \ \ \ \ \ \ \ \ \ \ \ \ \ \ \ \ \ \times\binom{\phi-1}{i-1}p_{S|S}^{\phi-i}p_{C|S}^{i-1}\Bigg].
\label{36} 
\end{align}

Conditioning on that SU transmission starts with $C$, the first two intervals are in state $C$ and $W$, respectively. We have
\begin{align}
&\Pr[T_{\textrm{ED}|C}=LT_s]=\notag\\
&\sum_{i=1}^{\phi}\sum_{\begin{subarray}{c}m,\ j\ \text{s.t.}\\m+2j=L-2-\phi-2(i-1)\end{subarray}}\Bigg[\binom{m+i-1}{i-1,\ j}p_{S|W}^ip_{C|W}^jp_{W|W}^{m-j}\notag\\
&\ \ \ \ \ \ \ \ \ \ \ \ \ \ \ \ \ \ \ \ \ \ \ \ \ \ \ \ \ \ \ \ \ \ \times\binom{\phi-1}{i-1}p_{S|S}^{\phi-i}p_{C|S}^{i-1}\Bigg].
\label{37}
\end{align}

Conditioning on starting with $S$, the differences are $N_{S|W}=N_{C|S}=i,\ i=1,2,\cdots,\phi-1$ and $N_{S|W}+N_{S|S}=\phi-1$. $\Pr[T_{\textrm{ED}|S}=LT_s]$ can be calculated as
\begin{align}
&\Pr[T_{\textrm{ED}|S}=LT_s]=\notag\\
&\sum_{i=1}^{\phi-1}\sum_{\begin{subarray}{c}m,\ j\ \text{s.t.}\\m+2j=L-\phi-2i\end{subarray}}\Bigg[\binom{m+i-1}{i-1,\ j}p_{S|W}^ip_{C|W}^jp_{W|W}^{m-j}\notag\\
&\ \ \ \ \ \ \ \ \ \ \ \ \ \ \ \ \ \times\binom{\phi-1}{i}p_{S|S}^{\phi-1-i}p_{C|S}^{i}\Bigg]+\delta(\phi)p_{S|S}^{\phi-1},
\label{38} 
\end{align}
where $\delta(\cdot)$ represents impulse function. $\delta(\phi)p_{S|S}^{\phi-1}$ is the special case that $T_{\textrm{ED}}$ is just a sequence of $\phi$ transmission intervals without collision. The final PMF of $T_{\textrm{ED}}$ is obtained by substituting Eq. (\ref{36}), Eq. (\ref{37}) and Eq. (\ref{38}) into Eq. (\ref{Ted_formulation_PS}).

\bibliographystyle{IEEEtran}
\bibliography{IEEEabrv,IEEETED}

\end{document}